\documentclass[a4paper,12pt]{article}
\usepackage{amssymb,amsmath,amsfonts,bbm}
\usepackage{graphicx}
\usepackage{mathrsfs}
\usepackage{multirow}
\numberwithin{equation}{section}

\begin{document}

\noindent
{\Large  {\bf Dependence function for bivariate cdf's}}\\

\vspace{0.5cm}
\noindent
{\Large {Teresa Ledwina}}\\


\vspace{1cm}

\noindent
ABSTRACT. Measuring a strength of dependence of  random variables is an important problem in statistical practice. In this paper, we propose a new function valued measure of dependence of two random variables. It allows one to study and visualize explicit dependence structure, both in some theoretical models and empirically, without prior model structure. This provides a comprehensive view of association structure and makes possible much detailed inference than based on standard numeric measures of association. We present theoretical properties of the new measure of dependence and discuss in detail estimation and application of copula-based variant of it. Some artificial and real data examples illustrate the behavior and practical utility of the measure and its estimator. \\

\vspace{1cm} 


\vspace{1cm}

\vspace{1cm}
\noindent
Teresa Ledwina, Institute of Mathematics, Polish Academy of Sciences, ul. Kopernika 18, 51-617 Wroc{\l}aw, Poland (
E-mail: {\it ledwina@impan.pan.wroc.pl}).
Research was supported by the Grant N N201 608440 from the National Science Center, Poland. The author thanks Dr. Grzegorz Wy{\l}upek for his help in preparing figures and tables of this article and for useful remarks.\\

\newpage
\vspace{1cm}
\noindent
{\bf 1. Introduction}\\

Measuring a strength of dependence of two random variables has long history and wide applications. For brief overview see Jogdeo (1982) and Lancaster (1982). More detailed information can be found in Drouet Mari and Kotz (2001) as well as Balakrishnan and Lai (2009), for example.  Most of measures of dependence, introduced in vast literature on the subject, are scalar ones. Such indices are called global measures of dependence. However, nowadays there is strong evidence that an attempt to represent complex dependence structure via a single number can be misleading. To overcome this drawback, several local indices have been proposed; see Section 6.3 of Drouet Mari and Kotz (2001) for details. Many of these indices were introduced in the context of regression models or survival analysis. Some local dependence functions have been introduced as well. In particular, Kowalczyk and Pleszczy\'nska (1977) invented function valued measure of monotonic dependence, based on some conditional expectations and adjusted to detect dependence weaker than the quadrant one. Next, Bjerve and Doksum (1993), Bairamov et al. (2003) and Li et al. (2014), among others,  introduced local dependence measures based on regression concepts. See the last mentioned paper for more information. Holland and Wang (1987) defined the local dependence function, which mimics cross-product ratios for bivariate densities and treats the two variables in a symmetrical way. This function valued measure has several appealing properties and received  considerable attention in the literature; cf. Jones and Koch (2003) for discussion and references. However, on the other hand, this measure has some limitations: it is not normalized, requires existence of densities of the bivariate distribution, and is intimately linked to strong form of dependence, the likelihood ratio dependence. Recently, Tj{\o}stheim and Hufthammer (2013) extensively discussed the role and history of local dependence measures in finance and econometrics. They also proposed the new local dependence measure, the local correlation function, based on approximating of bivariate density locally by a family of Gaussian densities. Similarly as the measure of Holland and Wang (1987), this measure treats both variables on the same basis.
Though the idea behind the construction of this measure is intuitive one its computation and estimation is a difficult and complex problem. 
The asymptotic theory developed in Tj{\o}stheim and Hufthammer (2013) treats in detail the problems mentioned above, in a scope that covers some time series models. In Berentsen et al. (2013) this theory is applied to describe dependence structure of different copula models. 

In this paper, we propose the new function valued measure of dependence of two random variables $X$ and $Y$ and present its properties. The measure has simple form and its definition exploits cumulative distribution functions (cdf's), only. In particular, we do not assume existence of a density of the observed vector. The measure takes values in [-1,1] and treats both variables in a symmetrical way. 
 The measure preserves the correlation order, or equivalently the concordance order, which is the quadrant order restricted to the class of distributions with fixed marginals. In particular, it is non-negative (non-positive) if and only if $X$ and $Y$ are positively (negatively) quadrant dependent. Quadrant dependence is relatively weak, intuitive and useful dependence notion, widely used in insurance and economics; see Dhaene et al. (2009) for an evidence and further references. 
The new measure obeys several properties formulated in the literature as useful or desirable. 
We introduce two variants of the measure. In Section 2 we consider general case, assuming that the vector $(X,Y)$ has joint cdf $H$ and marginals $F$ and $G$, respectively.  In Section 3 we discuss its copula-based counterpart which corresponds to some cdf $C$ on $[0,1]^2$ with uniform marginals.
Both variants allows for readable visualization of departures from independence. We focus our presentation on the copula-based variant.
Simple and natural estimator of the copula-based measure in the i.i.d. case is proposed and its appealing properties are discussed. The estimator can be effectively exploited to assess graphically underlying bivariate dependence structure and to build some formal local and global tests. Some illustrative examples are given in Section 4 to support utility of new solution. Section 6 concludes. \\

\noindent
{\bf 2. General case}\\

Consider a pair of random variables $X$ and $Y$ with cdf's $F(x)=P(X\leq x)$ and $G(y)=P(Y\leq y)$, respectively and a joint cdf $H(x,y)=P(X\leq x,Y\leq y)$.  Set $\mathbbm{D} =\{(x,y): 0<F(x)<1,0<G(y)<1\}$ and define
$$
q(x,y)=q_H(x,y)=\frac{H(x,y)-F(x)G(y)}{\sqrt{F(x)G(y)[1-F(x)][1-G(y)]}}\;\;\; \mbox{for}\;\;(x,y)\in \mathbbm{D}.
\eqno(1)
$$
From (1) it is seen that $q$ treats both variables $X$ and $Y$ symmetrically and a knowledge of $q$ and the marginal distributions allows one to recover $H$. The measure $q$ fulfills the following  properties, motivated by the axioms formulated in Schweitzer and Wolff (1981) and updated in Embrechts et al. (2002). \\

\noindent
{\bf Proposition 1.}\\

\noindent
1. $q$ is defined for any $X$ and $Y$.\\
2. $-1 \leq q \leq 1$ for all $(x,y) \in \mathbbm{D}$.\\
3. If the variables $X$ and $Y$ are exchangeable then $q(x,y)=q(y,x)$ for $(x,y) \in \mathbbm{D}$.\\
4. $q(x,y)\equiv 0$ if and only if $X$ and $Y$ are independent.\\
5. $q$ is non-negative (non-positive) if and only if $(X,Y)$ are positively (negatively) quadrant dependent.\\
6. $q$ is maximal (minimal) if and only if $Y=f(X)$ and $f$ is non-decreasing (non-increasing) a.s. on the range of $X$.\\
7. $q$ respects concordance ordering, i.e. for cdf's $H_1$ and $H_2$ with the same marginals, $H_1(x,y) \leq H_2(x,y)$ for all $(x,y) \in \mathbbm{R}^2$ implies $q_{H_1}(x,y) \leq q_{H_2}(x,y)$ for all $(x,y) \in \mathbbm{D}$.\\

\noindent
{\bf Proof}. Most of the above mentioned properties are obvious. The property 6 is an immediate consequence of Fr\'echet-Hoeffding bounds and their properties. To justify 2 it is enough to show that $q_H(x,y)$ is the correlation coefficient of some random variables. For this purpose, for $(x,y) \in \mathbbm{D}$ and 
$(s,t) \in \mathbbm{R}^2$ set
$$
\phi_x(s)=-\sqrt{\frac{1-F(x)}{F(x)}}\mathbbm{1}_{(-\infty ,x]}(s) + \sqrt{\frac{F(x)}{1-F(x)}}\mathbbm{1}_{(x, +\infty)}(s),
$$
$$
\psi_y(t)=-\sqrt{\frac{1-G(y)}{G(y)}}\mathbbm{1}_{(-\infty ,y]}(t) + \sqrt{\frac{G(y)}{1-G(y)}}\mathbbm{1}_{(y, +\infty)}(t).
$$
Then, by an elementary argument one gets
$$
q(x,y)=q_H(x,y)=E_H\phi_x(X)\psi_y(Y)=Cov_H\phi_x(X)\psi_y(Y)=Corr_H\phi_x(X)\psi_y(Y).
\eqno(2)
$$
\hfill \quad $\square$\\

\noindent
{\bf Remark 1.} The last expression in (2) shows that the function $q$ is based on  aggregated local correlations. Moreover, note that $\int_{\mathbbm{R}}\phi_x(s)dF(s)=\int_{\mathbbm{R}}\psi_y(t)dG(t)=0 $ and $\int_{\mathbbm{R}}\phi_x^2(s)dF(s)=\int_{\mathbbm{R}}\psi_y^2(t)dG(t)=1 $, for all $(x,y) \in \mathbbm{D}$. Therefore, the value $q(x,y)$ can be interpreted as the Fourier coefficient of the cdf $H(s,t)$ pertaining to the quasi-monotone function $\phi_x(s) \psi_y(t),\;(s,t) \in \mathbbm{R}^2$.\\

Given random sample $(X_1,Y_1),\ldots,(X_n,Y_n)$ from cdf $H$ with marginals $F$ and $G$, set $H_n$, $F_n$, and $G_n$ for respective empirical cdf's. A natural estimator $\hat {q}_H$ of $q_H(x,y)$ results by plugging these empirical cdf's into (1). This, given $(x,y)$, yields rank statistics. Note that the values $\chi_{ni}$ given by $\hat {q}_H(X_i,Y_i),\; i=1,\ldots,n$, have been already introduced in Fisher and Switzer (1985), as one of the two components of the so-called chi-plots, designed to investigate possible patterns of association of two random variables. 

We shall not study the estimator $\hat {q}_H$ in this paper. In the next section we comment in more detail on special case of (1), and the related problems, in the case  when the role of $H$ is played by the pertaining  copula.\\

\noindent
{\bf 3. Copula-based measure of dependence}\\

In this  section, to avoid technicalities and to concentrate on the main idea, we restrict attention to cdf's $H$ with continuous marginals $F$ and $G$. Under such a restriction there exists a unique copula $C$ such that $H(x,y)=C(F(x),G(y))$. In other words, $C$ is the restriction to the unit square of the joint cdf of $(F(X),G(Y))$. The copula captures the dependence structure among $X$ and $Y$, irrespective of their marginal cdf's. This is important in many applications. For the related discussion see P\'oczos et al. (2012).\\

\noindent
{\sl 3.1. The form and further properties of q}\\

We have
$$
q(u,v)=q_{C}(u,v)=\frac{C(u,v)-uv}{\sqrt{uv(1-u)(1-v)}} = w(u,v)[C(u,v)-uv], \;\;\;\;(u,v) \in (0,1)^2,
\eqno(3)
$$
where
$$
w(u,v)=\frac{1}{\sqrt{uv(1-u)(1-v)}}.
\eqno(4)
$$
The interpretation of $q$ in terms of correlations, given in (2), is still valid with some obvious adjustment. Namely, now for $u \in (0,1)$ and $s\in (0,1)$ we consider
$$
\phi_u(s)=\psi_u(s)=-\sqrt{\frac{1-u}{u}}\mathbbm{1}_{[0,u]}(s)+\sqrt{\frac{u}{1-u}}\mathbbm{1}_{(u,1]}(s).
$$ 

\noindent
{\bf Proposition 2.} The copula based measure of dependence $q$, given by (3), additionally to 1-7, has the following properties.\\

\noindent
8. $q$ is invariant to strictly increasing a.s. on ranges of $X$ and $Y$, respectively, transformations.\\
9. If $X$ and $Y$ are transformed by strictly decreasing a.s. functions then $q(u,v)$ transforms to $q(1-u,1-v)$.\\
10. If $f$ and $g$ are strictly decreasing a.s. on ranges of $X$ and $Y$, respectively, then $q$'s for the pairs $(f(x),Y)$ and $(X,g(Y))$ take the forms $-q(1-u,v)$ and $-q(u,1-v)$, accordingly.\\
11. The equation $q(u,v)\equiv c$, $c$ a constant, can hold true if and only if $c=0$.\\
12. If $(X,Y)$ and $(X_n,Y_n),\;n=1,2,\ldots,$ are pairs of random variables with joint cdf's $H$ and $H_n$, and the pertaining copulas $C$ and $C_n$, respectively, then weak convergence of $\{H_n\}$ to $H$ implies $q_{C_n}(u,v) \to q_C(u,v)$ for each $(u,v) \in (0,1)^2$.\\

\noindent
{\bf Proof.} Properties 8-10 follow from Theorem 3 in Schweizer and Wolff (1981). The convergence in 12 is due to continuity of $C$. To justify 11 observe that the equation is equivalent to $C(u,v)=C_c(u,v)=uv+c\sqrt{uv(1-u)(1-v)}$. Since $C$ is quasi-monotone, then $C_c(u,v)$ should also possess such a property. Since $C_c(u,v)$ is absolutely continuous then quasi-monotonicity is equivalent to $\frac{\partial^2}{\partial u \partial v} C_c(u,v) \geq 0$ for almost all $(u,v) \in [0,1]$ (in the Lebesgue measure); cf. Cambanis et al. (1976). However, $\frac{\partial^2}{\partial u \partial v} C_c(u,v) = 1 + c[u-1/2][v-1/2]w(u,v)$ and for $c\not= 0$ this expression can be negative on the set of positive Lebesgue measure.\hfill \quad $\square$\\

\noindent
{\bf Remark 2.} The properties 4 and 8-10 provide some compromise to too demanding postulates P4 and P5 discussed in Embrechts et al. (2002). \\

\noindent
{\bf Remark 3.} The property 11 is very different from respective property of the local dependence function of Holland and Wang (1987) which is constant for the bivariate normal distribution and some other models; cf. Jones (1998) for details.\\

\noindent
{\bf Remark 4.} By Fr\'echet-Hoeffding bounds for copulas, the property 2 saying that $q(u,v) \in [-1,1]$ can be further sharpened to 
$$
B_*(u,v) \leq q_C(u,v) \leq B^*(u,v),\;\;\;(u,v) \in (0,1)^2
$$
where $B_*(u,v)=w(u,v)[\max\{u+v-1,0\} -uv]$ and $B^*(u,v)=w(u,v)[\min\{u,v\} -uv]$. Note that $B_*(u,1-u) = -1$ for $u \in (0,1)$ while $-1 < B_*(u,v) \leq 0$ otherwise. Similarly, $B^*(u,u)=+1$ for $u \in (0,1)$ and $0 \leq B^*(u,v)<1$ in the remaining cases. 
In Figure 1 we show these bounds on the the regular grid
$$
\mathbbm{G}_{16} = \{(u,v): u=i/16, v=j/16,\; i,j=1,\ldots,15\}.
$$

\noindent
\hspace{49 mm}
\includegraphics[width = 4 cm, height = 3.97 cm]{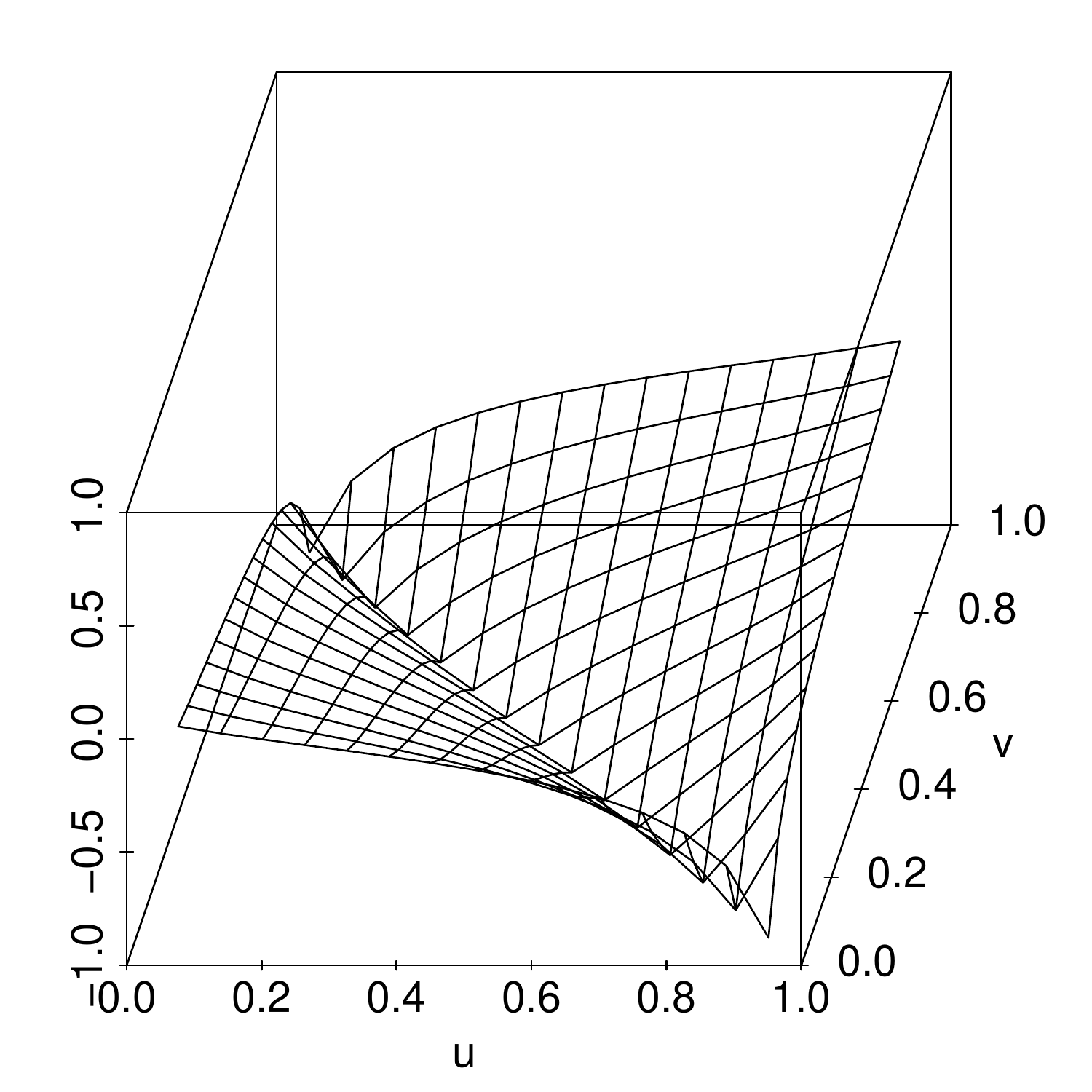}\quad 
\includegraphics[width = 4 cm, height = 3.97 cm]{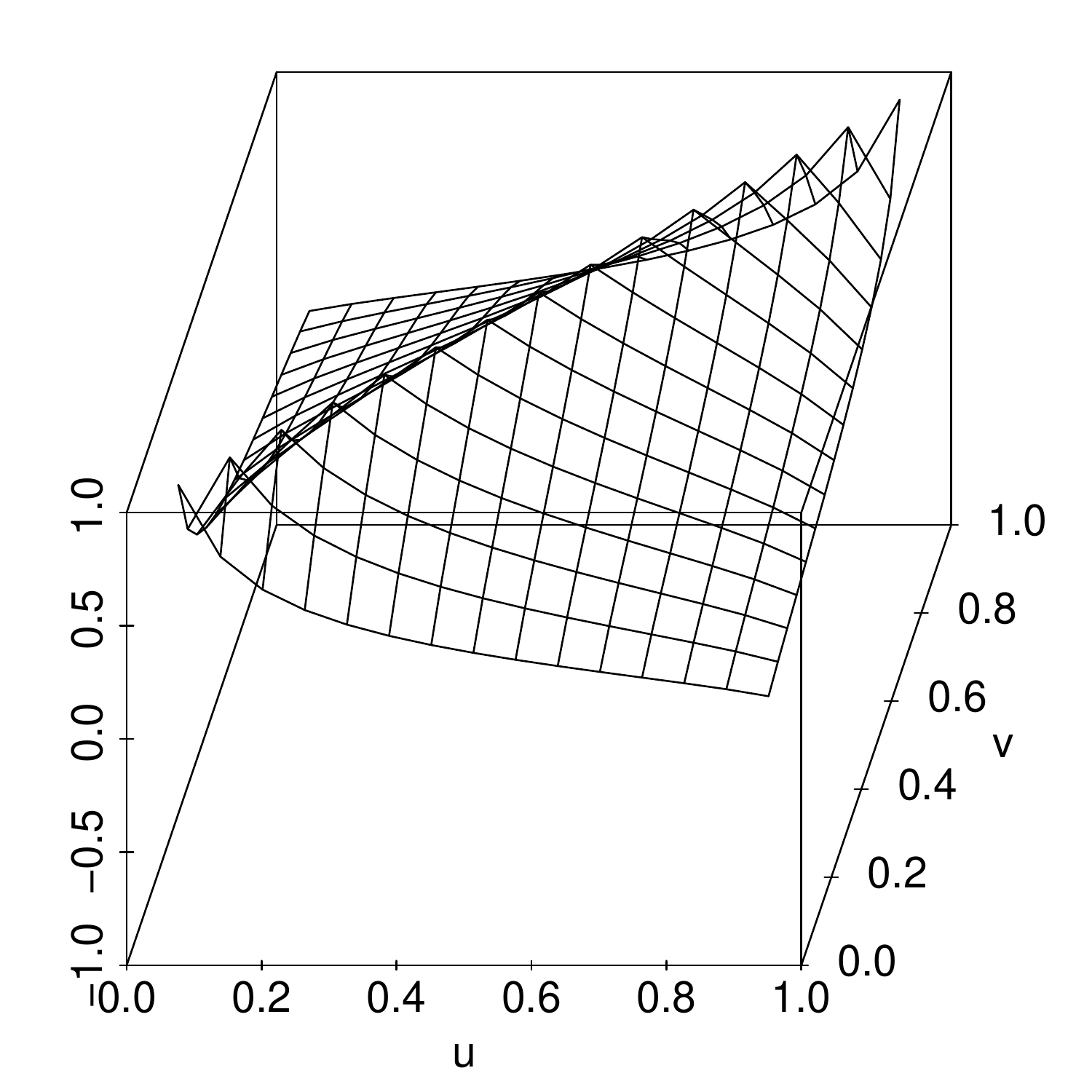}\quad

\noindent
{\it Fig. 1.} Left panel: lower bound $B_{*}(u, v)$ of $q_C(u, v)$ on the grid $\mathbbm {G}_{16}$;
right panel: upper bound $B^{*}(u, v)$ of $q_C(u, v)$ on the grid $\mathbbm {G}_{16}$.\\

\noindent
{\bf Remark 5.} Similarly as $B_*$ and $B^*$, the measure $q_C$ can be  displayed in a graphical form. This is useful feature helping to visualize a structure of departures from independence. It is also worth  emphasizing that though we focused of cdf's with continuous marginals we still do not assume that $H$ or $C$ posses a density.
In Figures 2 and 3 we show plots of $q_C$ pertaining to classical Marshall-Olkin and recently introduced Mai-Scherer (2011) extreme value copulas. The displays are accompanied by scatter plots of simulated data and pertaining heat maps. Therefore, we introduce first a natural estimator of $q_C$ and present all illustrative examples in Section 4.\\

\noindent
{\sl 3.2. Estimates of $q_C$}\\

Let $(X_1,Y_1),\ldots,(X_n,Y_n)$ be a random sample from cdf $H$. Furthermore, let $R_i$ be the rank of $X_i,\;i=1,\ldots,n$, in the sample $X_1,\ldots,X_n$ and $S_i$ the rank of $Y_i,\;i=1,\ldots,n,$ within $Y_1,\ldots,Y_n$. 
Simple estimate of $q_C$ has the form 
$$
\hat q_C (u,v)=w(u,v)[{D_n(u,v)-uv}], \;\;\;\;(u,v) \in (0,1)^2,
$$
where $D_n$ is rank-based empirical copula estimator of $C$, i.e.
$$
D_n(u,v)=\frac{1}{n}\sum_{i=1}^n \mathbbm{1}\Bigl(\frac{R_i}{n}\leq u,\frac{S_i}{n}\leq v\Bigr), \quad 
(u,v) \in [0,1]^2.
\eqno(5)
$$
The paper of Swanepoel and Allison (2013) provides exact mean and variance of $D_n(u,v)$. Ledwina and Wy{\l}upek (2014) have shown that $D_n(u,v)$ preserves the quadrant order. More precisely, if a copula $C_1$ has larger quadrant dependence than a copula $C_{2}$ then, under any fixed $(u,v) \in [0,1]^2$, any $c \in \mathbb{R}$ and any $n$, it holds that 
$$
P_{C_1}\Bigl(D_n(u,v) \geq c\Bigr) \geq P_{C_{2}}\Bigl(D_n(u,v)\geq c\Bigr).
\eqno(6)
$$
An obvious consequence of (6) is analogous order preserving property of $\hat q_C (u,v)$.

Asymptotic properties of the empirical copula process $Z_n(u,v)=\{\sqrt n [D_n(u,v)-uv],\;u,v \in [0,1]\}$ have been studied by many authors; cf. Fermanian et al. (2004) for the related results and references. Since $uv(1-u)(1-v)$ is the asymptotic variance of $Z_n(u,v)$ under independence, $\sqrt n \hat q_C (u,v)$  coincides with natural weighted empirical copula process while Theorem 3 of Fermanian et al. (2004) implies that, under independence, $\sqrt n \hat q_C (u,v)$ is asymptotically $N(0,1)$ for each $u,v \in (0,1)$. The estimate $D_n(u,v),$ in a series of papers originated by Deheuvels (1979), has been called empirical dependence function.  

In application oriented papers, more popular variant of rank-based estimator of $C$ is
$$
C_n(u,v)=\frac{1}{n}\sum_{i=1}^n \mathbbm{1}\Bigl(\frac{R_i}{n+1}\leq u,\frac{S_i}{n+1}\leq v\Bigr), \quad 
(u,v) \in [0,1]^2.
\eqno(7)
$$
The variables $(R_i/(n+1),S_i/(n+1)),\; i=1,\ldots,n$ are called pseudo-observations in the literature. Obviously, the finite sample and basic asymptotic properties of $C_n(u,v)$ are inherited after $D_n(u,v)$. Therefore, we shall consider the following estimator of $q_C$
$$
Q_n(u,v)=w(u,v)[{C_n(u,v)-uv}] = \frac{C_n(u,v)-uv}{\sqrt{uv(1-u)(1-v)}}, \;\;\;\;(u,v) \in (0,1)^2,
\eqno(8)
$$
which, similarly as $\hat q_C (u,v)$, is the rank statistic.
Moreover, we set 
$$
L_n(u,v)=\sqrt n Q_n(u,v)
\eqno(9)
$$
for the standardized version of this estimate. Simple algebra yields that for any $(u,v) \in (0,1)^2$ it holds
$$
L_n(u,v)=\frac{1}{\sqrt n} \sum_{i=1}^n\phi_u(\frac{R_i}{n+1})\phi_v(\frac{S_i}{n+1}) + O(\frac{1}{\sqrt n}).
\eqno(10)
$$
So, up to deterministic term of the order $O(1/\sqrt n)$, the standardized estimator $L_n(u,v)$ is linear rank statistic with the quasi-monotone score generating function $\phi_u \times \phi_v$. Moreover, the definition of $L_n$ and (6) yield that 
$$
P_{C_1}\Bigl(L_n(u,v) \geq c\Bigr) \geq P_{C_{2}}\Bigl(L_n(u,v)\geq c\Bigr)
\eqno(11)
$$
for any $(u,v) \in (0,1)^2$, any $c$, any $n$, and any two copulas $C_1$  and $C_2$ such that $C_1$ has larger quadrant dependence than $C_2$.
Summarizing the above mentioned results, let us note that under independence $L_n(u,v)$ is distribution free. So, given $n$, under independence, the significance of the obtained values of this statistic can be easily assessed on a basis of simple simulation experiment. For large $n$ one can rely on asymptotic normality of $L_n(u,v)$. Due to (11), similar conclusions follow if one likes to verify hypothesis asserting that $q_C(u,v) \geq 0.$ Moreover, (11) implies that different levels of strength of quadrant dependence of the underlying $H$'s shall be adequately quantified by order preserving $L_n(u,v)$'s.
These results make the values of  $L_n(u,v), \;(u,v) \in (0,1)^2,$ a useful diagnostic tool. For example, since quadrant dependence is relatively weak notion, significantly negative values of  $L_n(u,v)$ for some $(u,v)$'s make questionable positive quadrant dependence, and many other forms of positive dependence of the data at hand, as well.

To close, note that, given $u$ and $u$,  the score generating function $\phi_u \times \phi_v$, appearing in (10), is not smooth one and takes on at most four possible values, only. This causes that, under independence, the convergence of $L_n(u,v)$ to the limiting $N(0,1)$ law is not very fast. Moreover, the rate of convergence is expected to depend on $u$ and $v$, with the least favorable situation when $(u,v)$ is close to the vertices of the unit square. We illustrate these aspects in Table 1, where simulated critical values of the test rejecting independence for large values of $|L_n(u,v)|$ are given under some $(u,v)$'s, five different sample sizes, and two selected significance levels $\alpha$. In cases when finite sample distribution of $L_n$ is far from continuous one, the problem of uniqueness of sample quantiles arises. Through, to calculate sample quantiles we apply Gumbel's approach, described in Hyndman and  Fan (1996) by Definition 7.\\

\noindent
Table 1. Simulated critical values of the test rejecting independence for large values of $|L_n(u,v)|$ for selected $(u,v)$, versus $n$ and $\alpha$.\\

 \begin{tabular}{|c||r|r|r|r|r|r|r|r|r|r|}\hline
 & \multicolumn{5}{c|}{$\alpha = 0.01$} & \multicolumn{5}{c|}{$\alpha = 0.05$} \\
\hline
$(u,v)$ & \multicolumn{5}{c|}{$n$} & \multicolumn{5}{c|}{$n$} \\
\hline
 & 200 & 300 & 400 & 500 & 600 & 200 & 300 & 400 & 500 & 600\\\hline
$(\frac{1}{2},\frac{1}{2})$     & 2.546 & 2.540 & 2.600 & 2.504 & 2.613 & 1.980 & 1.848 & 2.000 & 1.968 & 1.960\\
$(\frac{1}{12},\frac{1}{12})$ & 2.520 & 2.960 & 2.800 & 2.716 & 2.583 & 1.594 & 1.575 & 1.782 & 1.968 & 2.049\\
$(\frac{1}{16},\frac{1}{16})$ & 2.753 & 2.879 & 2.933 & 2.349 & 2.591 & 1.546 & 1.894 & 2.080 & 1.586 & 1.894\\
$(\frac{1}{20},\frac{1}{20})$ & 2.233 & 2.735 & 3.158 & 2.589 & 3.008 & 2.233 & 1.519 & 2.105 & 1.648 & 2.149\\ \hline
\end{tabular}\\

\vspace{0.5cm}

The sample sizes in Table 1 are close to that we shall consider in Section 4. The results exhibit that, for this range of sample sizes, the simulated quantiles of $|L_n(1/2,1/2)|$ are reasonably close to the limiting ones. We studied some selection of regular grids $\{(i/g,j/g),\;i,j=1,\ldots,g-1\}$, for some $g$'s, the related behavior of $L_n$'s and reported in Table 1 the results on $L_n(1/g,1/g)$ for $g=12, 16, 20$. They illustrate the conclusion that too dense grid shall imply too much inaccuracy in the simulated quantiles while too conservative choice not necessarily provides much progress. In view of these observations, we decided to apply the regular grid $\mathbbm{G}_{16}$ through. With this choice, for the sample sizes under consideration, the significance of observed values of $|L_n(i/g,j/g)|$ can be easily approximately evaluated looking at the heat maps, that we provide in each case. For more precise evaluation extra simulations are needed.

Note also that there are available some smooth nonparametric estimators of copulas. See Janssen et al. (2012) and Omelka et al. (2009) for recent contributions and extensive overview. However, we do prefer to insert the estimator $C_n$, since it is naturally linked to local correlations defined in the paper and obeys the property (11), which is crucial in finite sample inference on dependence. Counterparts of (11) for more general rank statistics are not available, according to the best our knowledge. For some related discussion see Ledwina and Wy{\l}upek (2014), Section 3.
\\

\noindent
{\bf 4. Illustration}\\

\noindent
{\sl 4.1. Example 1:  Extreme value copulas}\\

We start with two simulated data sets of size $n=500$ from Marshall-Olkin and Mai-Scherer (2001) copulas given by $C(u,v)=C_{a,b}^{MO}(u,v)=\min\{u^{1-a}v,uv^{1-b}\},\;a=0.50, b=0.75$, and $C(u,v)=C_{a,b}^{MS}(u,v)=\min\{u^a,v^b\}\min\{u^{1-a},v^{1-b}\},\;a=0.9, b=0.5$; cf. Nelsen (2006), p. 53 and Mai-Scherer (2001), p. 313, respectively. Both copulas possess a singular part. In Figures 2 and 3 we show dependence functions $q_C(u,v)$ for these models. The functions are accompanied by scatter plots of pseudo-observations $(R_i/(n+1),S_i/(n+1)), i=1,\ldots,500$, from the simulated samples. The scatter plots nicely exhibit the singularities and show some tendencies in the data. Right panels in these figures display respective heat maps of standardized correlations $L_n(u,v)$'s calculated on the grid 
$\mathbbm {G}_{16}$. Each square of size $0.0625\times0.0625$ represents the respective value of $L_n$ in its upper-right corner. To simplify reading, each heat map is accompanied with two numbers 
$$
L_* = \min_{1\leq i,j \leq 15} L_n(i/16,j/16) \;\;\;\mbox{and}\;\;\; L^*=\max_{1\leq i,j \leq 15} L_n(i/16,j/16).
\eqno(12)
$$
Both copulas represent positively quadrant dependent distributions. Under such dependence large values of $U$ tend to associate to large values of $V$ and similar pattern applies to small values. This tendency is nicely seen in the figures. Intuitively, the tendency is stronger for $C_{0.9,0.5}^{MS}$ than  for $C_{0.50,0.75}^{MO}$ and this is indeed well reflected by the heat maps. The points of the grid $\mathbbm{G}_{16}$ in which the estimated correlations $Q_n$ are significant on the levels 0.05 and 0.01 can be easily identified; cf. Table 1. Some possibility of testing for positive local and/or global dependence is sketched in Section 4.2.\\

\hspace{40 mm} Marshall-Olkin copula, $\alpha = 1/2$, $\beta = 3/4$\\

\vspace{1 mm}

\noindent
\includegraphics[width = 4 cm, height = 3.97 cm]{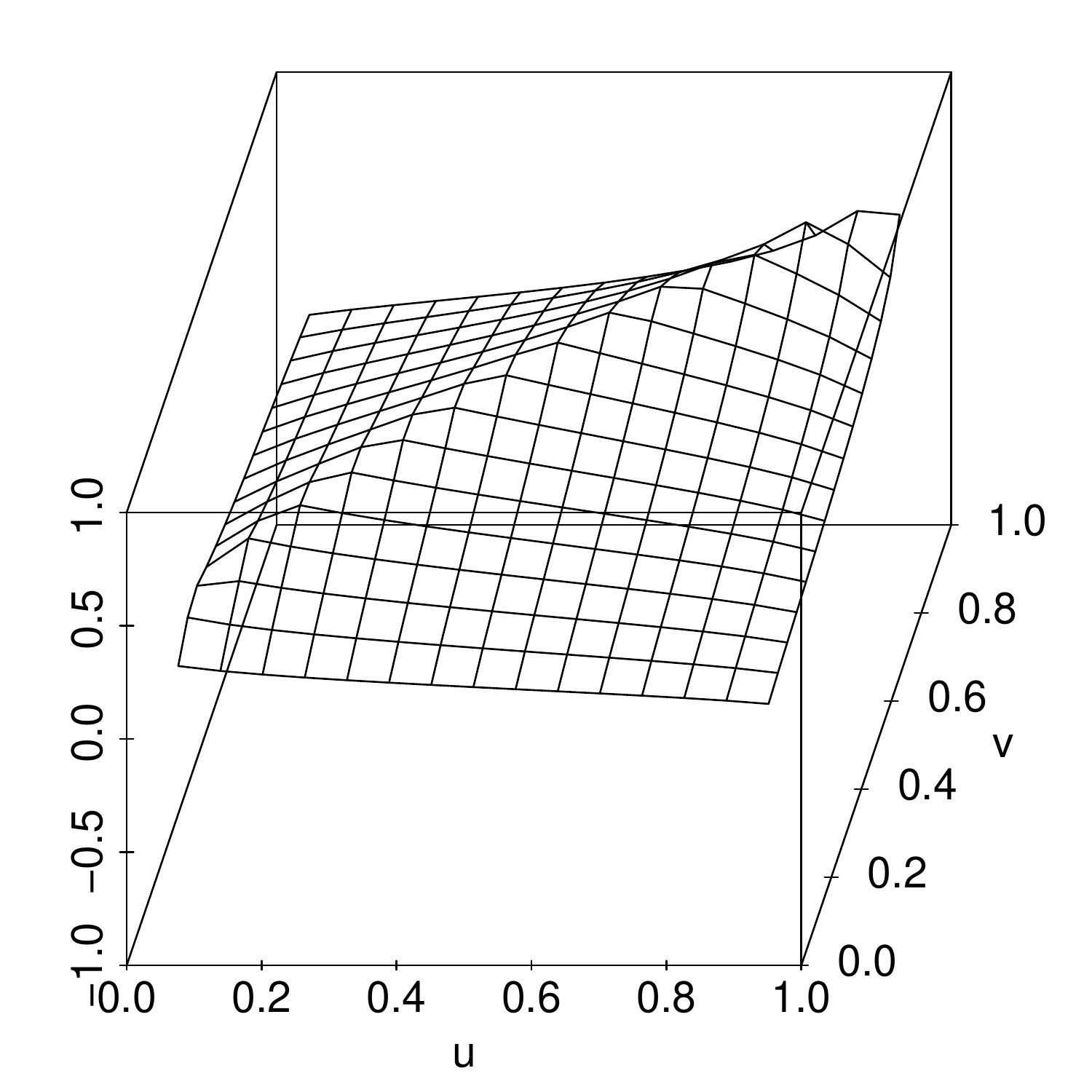}\quad
\includegraphics[width = 4 cm, height = 3.97 cm]{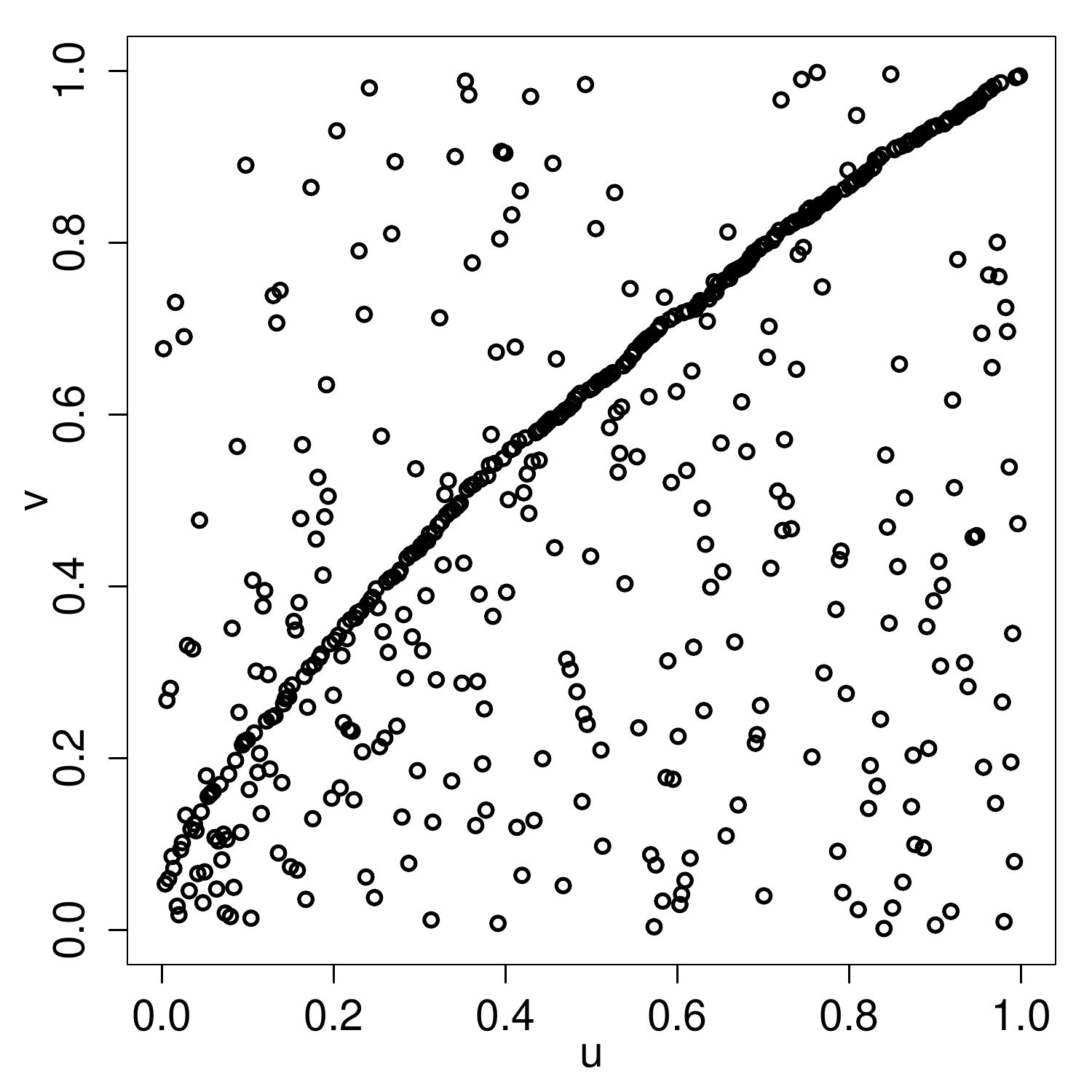}\quad\quad
\includegraphics[width = 4 cm, height = 3.97 cm]{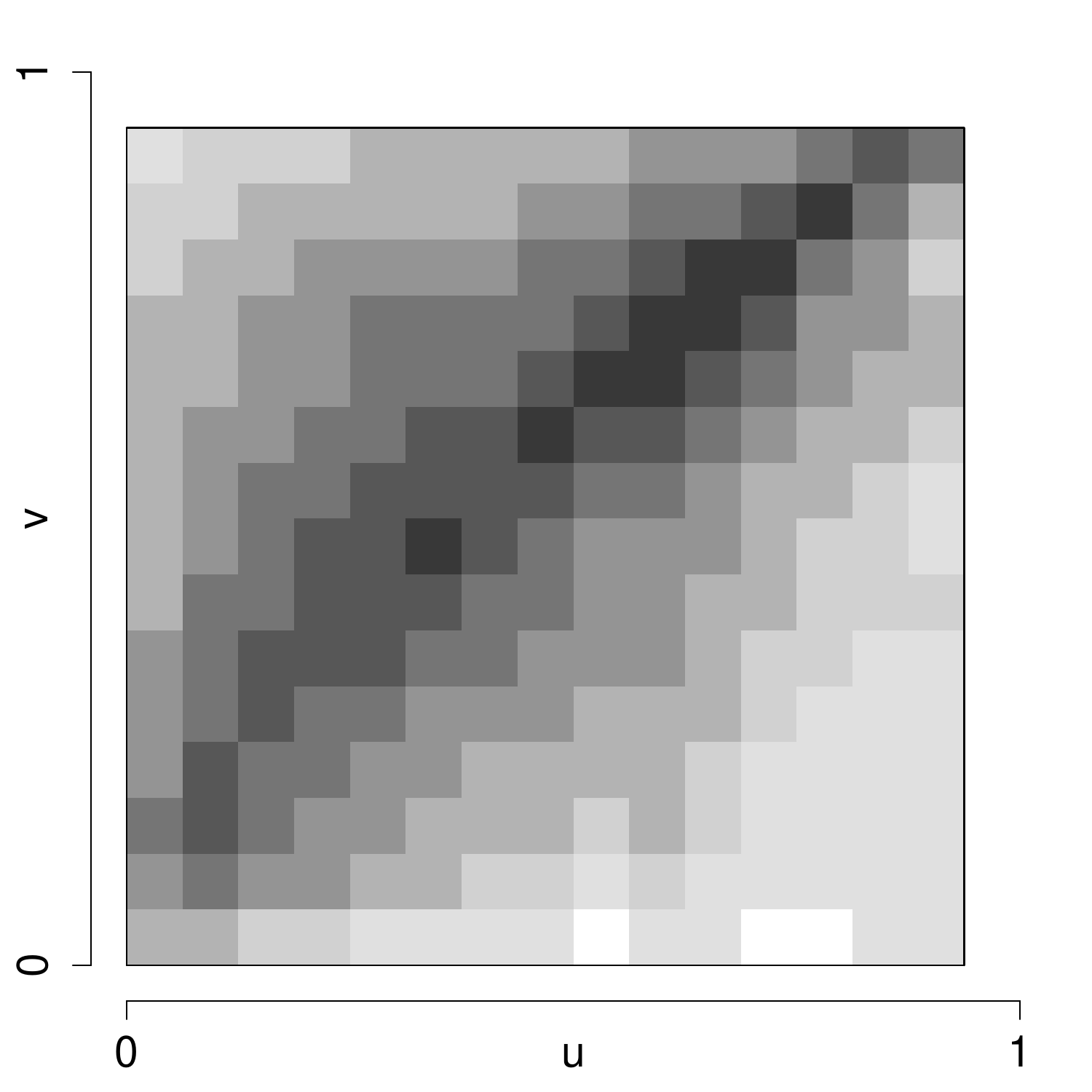}
\includegraphics[width = 4 cm, height = 3.97 cm]{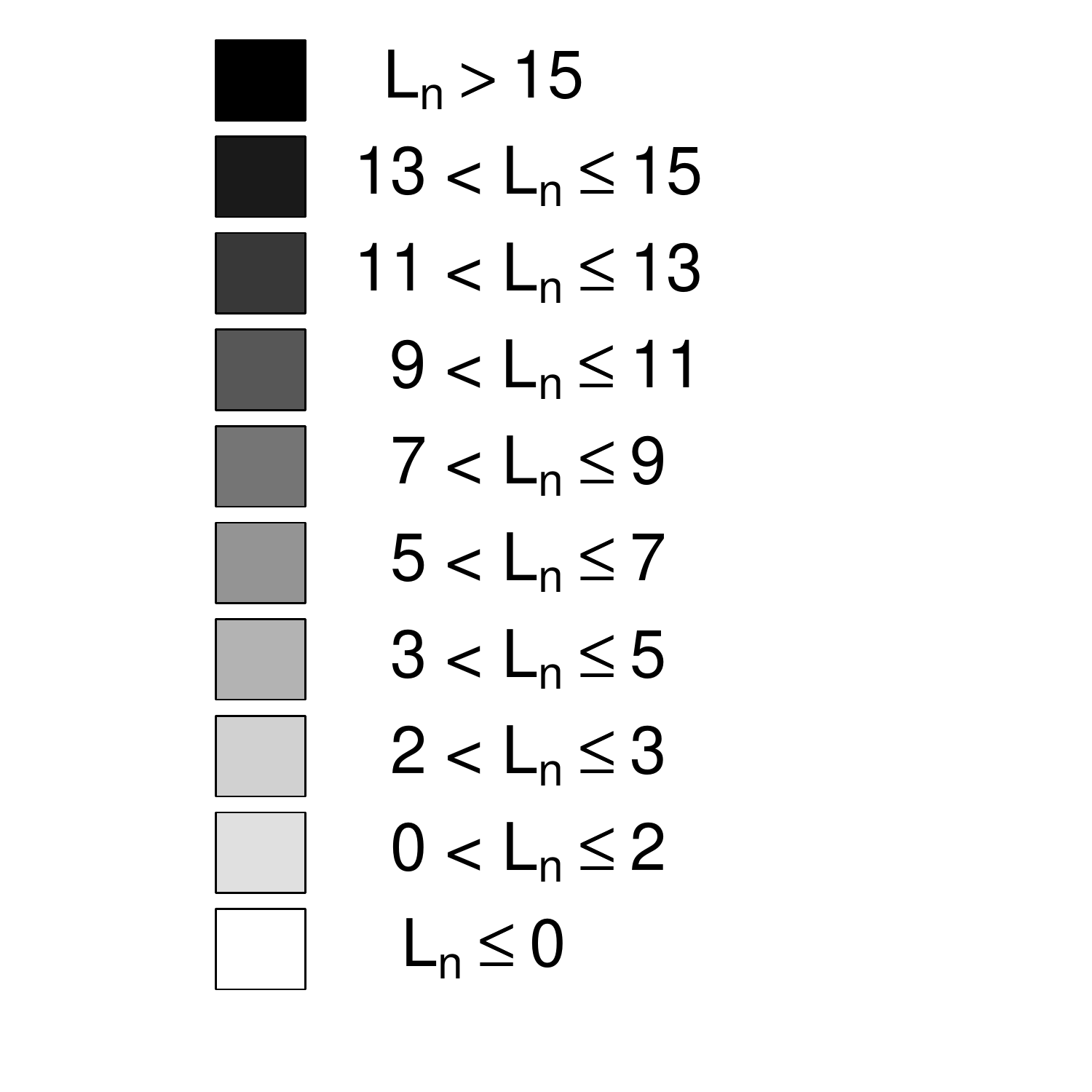}

\noindent
{\it Fig. 2.} Left panel: dependence function $q_C(u, v)$ for the Marshall-Olkin copula;
middle panel: scatter plot of \,$(R_i/(n+1),S_i/(n+1))$,\, $i = 1,\ldots,n$, $n = 500$,
of simulated observations from the copula; right panel: standardized estimator $L_n(u,v)$
of $q_C(u, v)$ on the grid $\mathbbm {G}_{16}$. $L_{*} = -0.2$, $L^{*} = 12.2$.\\

\newpage

\hspace{40 mm} Mai-Scherer copula, $a = 0.9$, $b = 0.5$\\

\vspace{1 mm}

\noindent
\includegraphics[width = 4 cm, height = 3.97 cm]{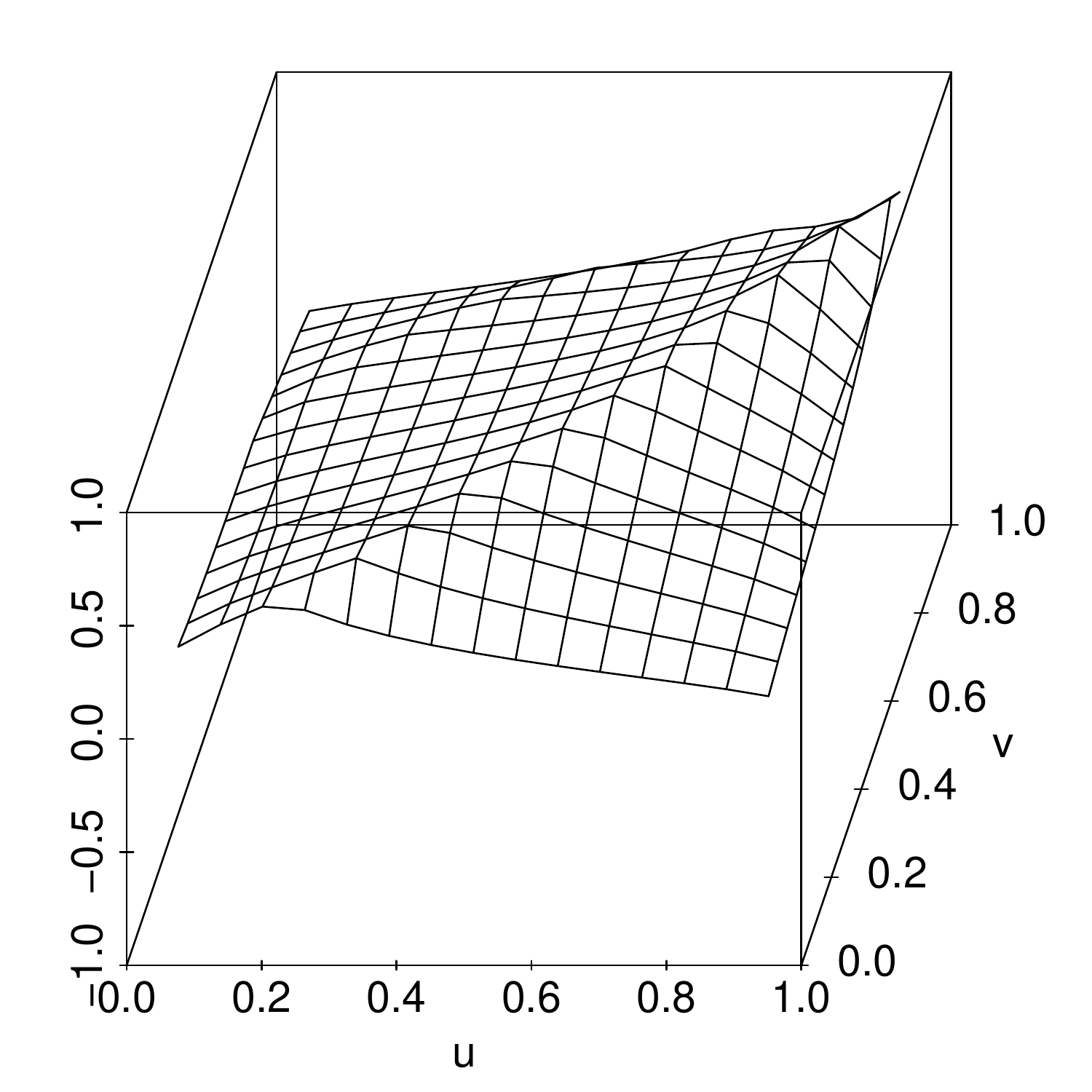}\quad
\includegraphics[width = 4 cm, height = 3.97 cm]{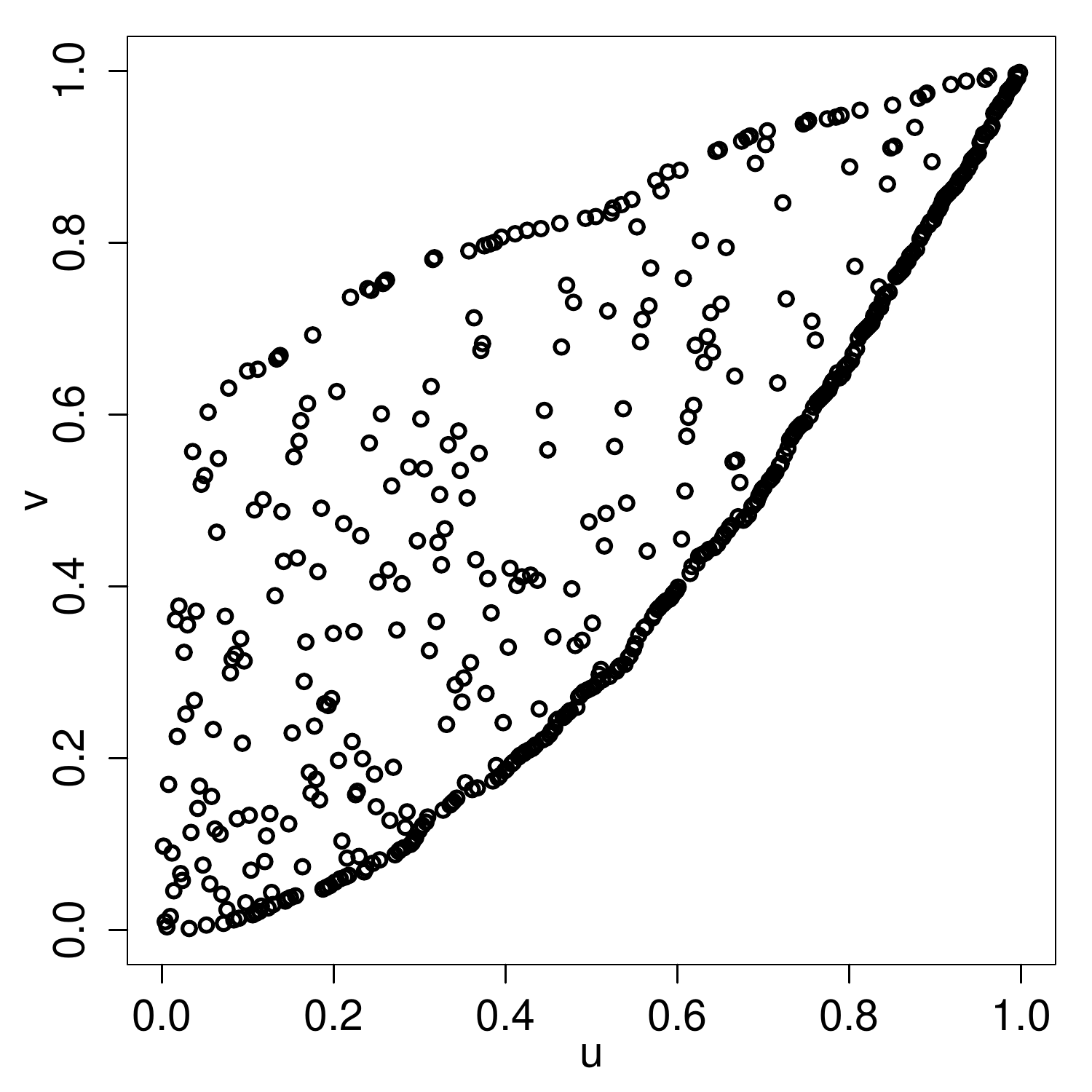}\quad\quad
\includegraphics[width = 4 cm, height = 3.97 cm]{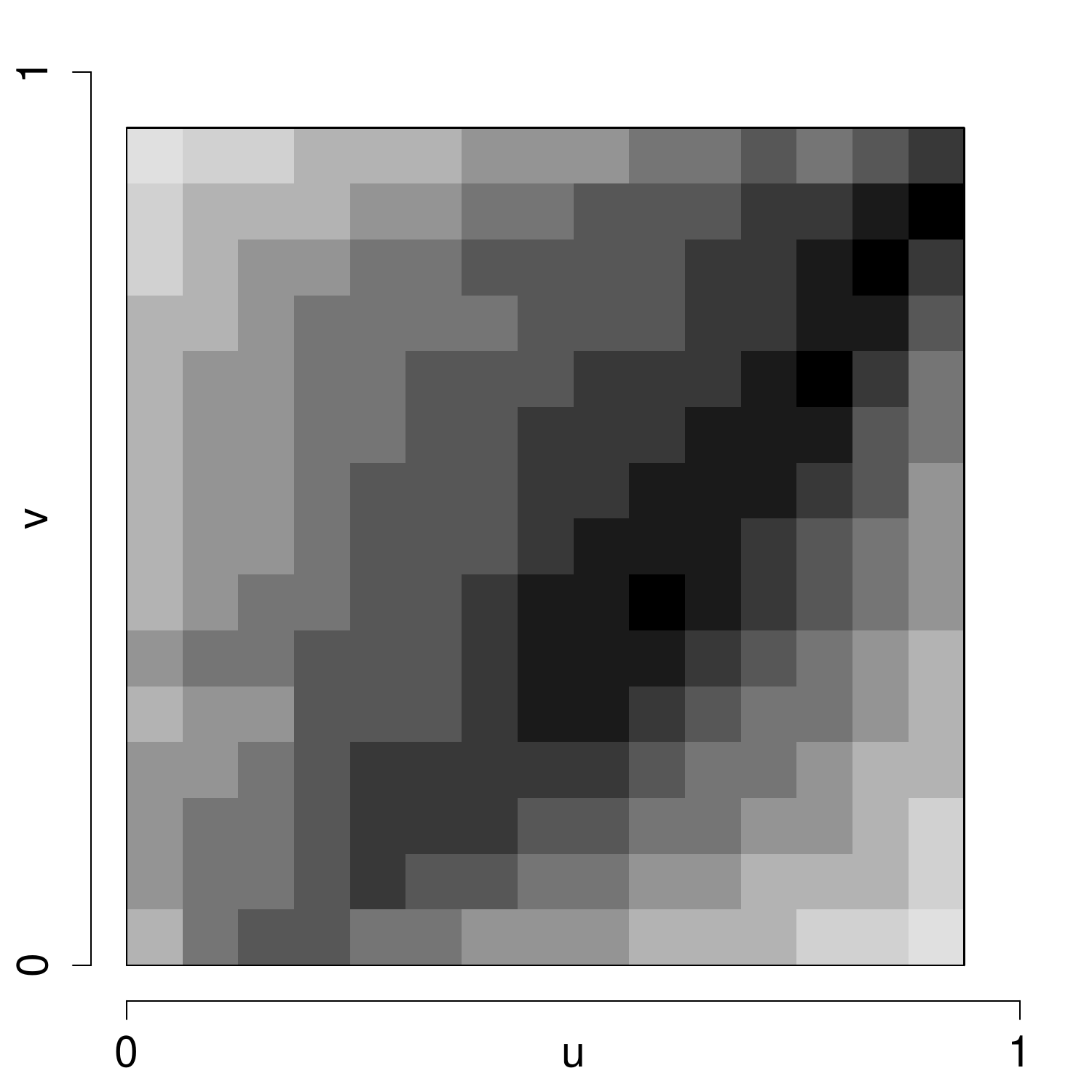}
\includegraphics[width = 4 cm, height = 3.97 cm]{Legend.pdf}

\noindent
{\it Fig. 3.} Left panel: dependence function $q_C(u, v)$ for the Mai-Scherer copula;
middle panel: scatter plot of \,$(R_i/(n+1),S_i/(n+1))$,\, $i = 1,\ldots,n$, $n = 500$,
of simulated observations from the copula; right panel: standardized estimator $L_n(u,v)$
of $q_C(u, v)$ on the grid $\mathbbm {G}_{16}$. $L_{*} = 1.5$, $L^{*} = 16.1$.\\

Next  examples follow similar pattern. They concern three real data sets considered earlier by Jones and Koch (2003). In each example of our paper we use the same scale of intensity of colors in the heat maps. This allows one to compare how different degrees of association are reflected by our estimators.\\

\noindent
{\sl 4.2. Example 2: Automobile data}\\

We shall consider two data sets of size $n=392$ available through 
{\tt www/http://lib.stat.\\cmu.edu/datasets/cars}. This is 1983 {\it ASA Data Exposition}  data set, collected by Ernesto Ramos and David Donoho.

The first sample collects observations of engine power (variable $X$), measured in horsepower, and fuel consumption (variable $Y$). 
This example was already investigated by Hawkins (1994), who fitted to the original data points decreasing regression function. Strong negative association is also clearly manifested by the scatter plot, which is based of transformed observations. \\

\hspace{40 mm} { (Engine Power, Fuel Consumption)}, $n = 392$\\

\vspace{1 mm}

\noindent
\includegraphics[width = 4 cm, height = 3.97 cm]{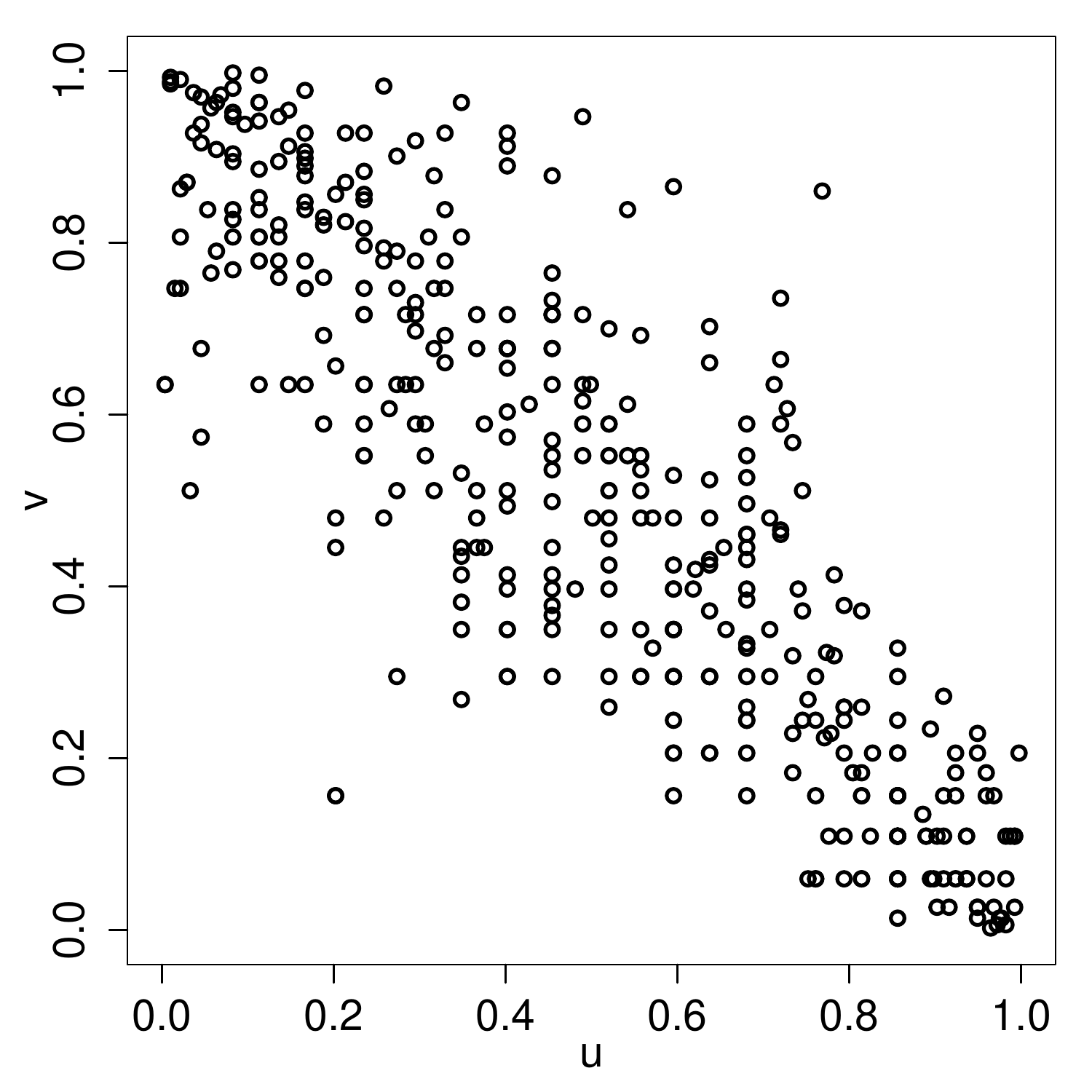}\quad \quad
\includegraphics[width = 4 cm, height = 3.97 cm]{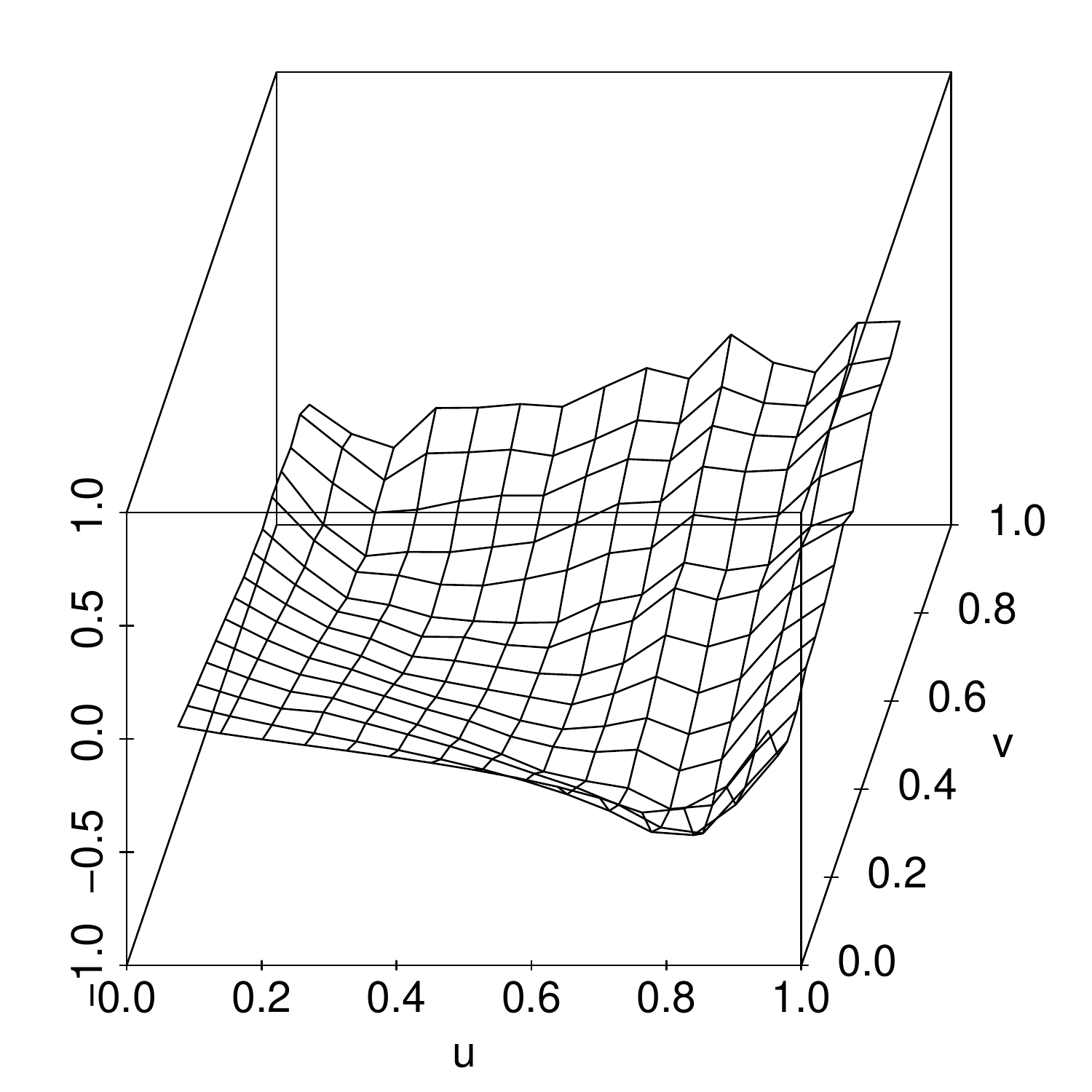}\quad
\includegraphics[width = 4 cm, height = 3.97 cm]{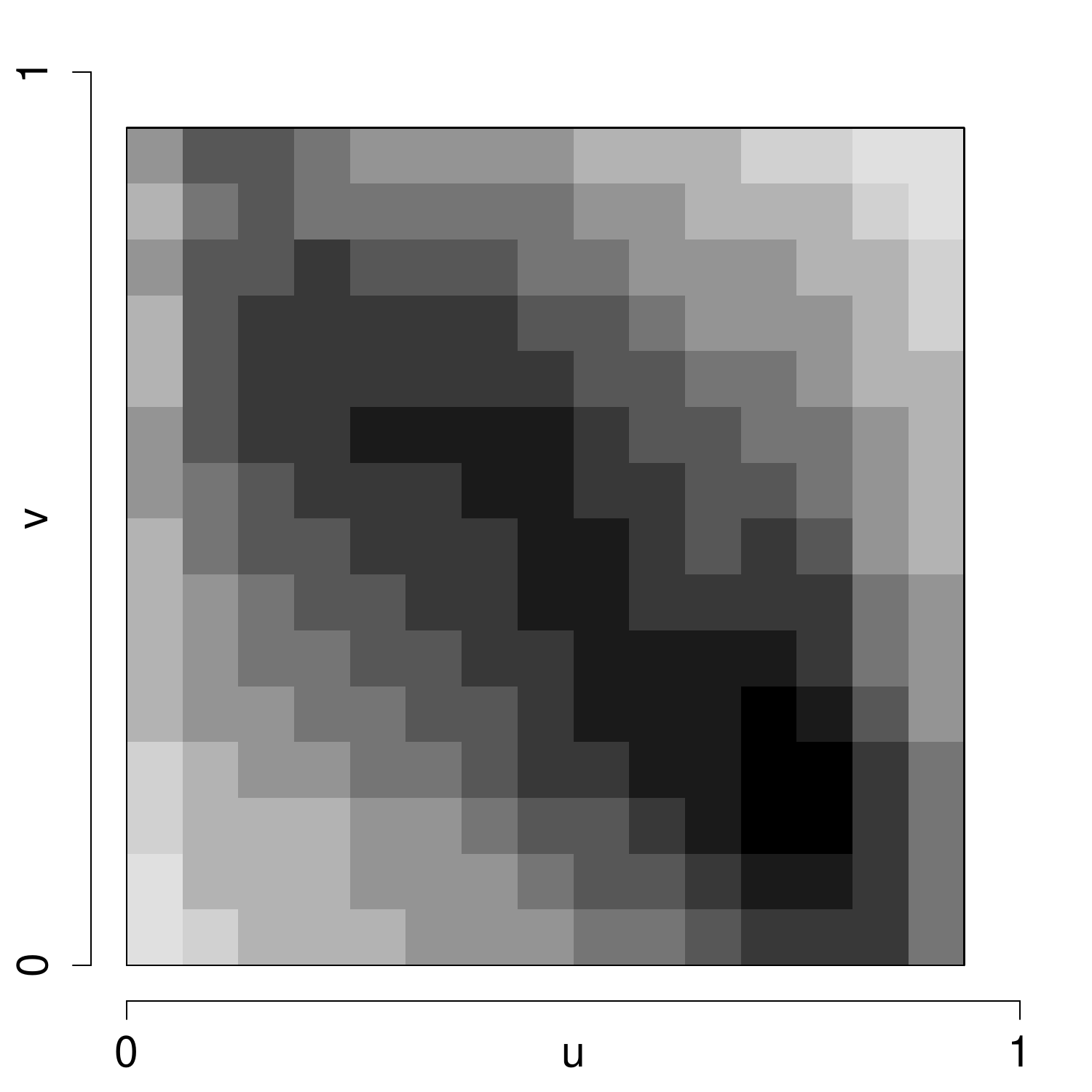}
\includegraphics[width = 4 cm, height = 3.97 cm]{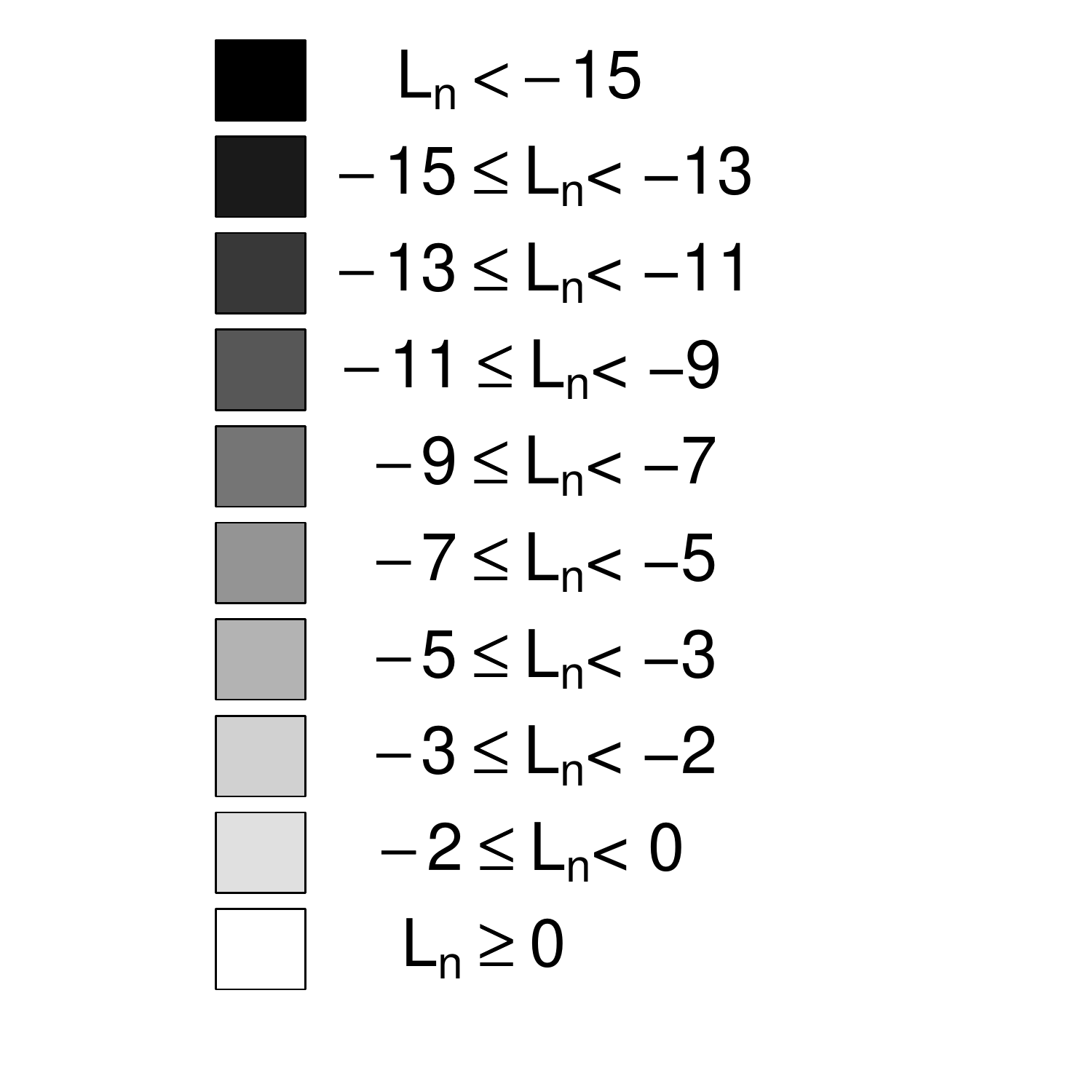}

\noindent
{\it Fig. 4.} Left panel: scatter plot of \,$(R_i/(n+1),S_i/(n+1))$,\, $i = 1,\ldots,n$, $n = 392$;
middle panel: estimator $Q_n(u,v)$ of $q_C(u, v)$ on the grid $\mathbbm {G}_{16}$;
right panel: standardized estimator $L_n(u,v) = \sqrt{n}\, Q_n(u,v)$ on the grid $\mathbbm {G}_{16}$.
$L_{*} = -16.0$, $L^{*} = -1.2$.\\

The heat map indicates visible negative trend and strong negative dependence in most of the points of the grid $\mathbbm {G}_{16}$. Only for some points close to the edges (0,0) and (1,1) the correlations are small in absolute value. To allow for some immediate quantitative analysis we give in Table 2 simulated quantiles of $L_n(u,v)$ for $n=392$ and two choices of $(u,v)$'s. \\

\noindent
Table 2. Simulated $\alpha$-quantiles  of $L_n(u,v)$ for two selected $(u,v)$ and $n=392$.\\

 \begin{tabular}{|c||r|r|r|r|r|r|}\hline
 & \multicolumn{6}{c|}{$\alpha $}  \\
\hline
 $(u,v)$ & 0.01 & 0.05& 0.10 & 0.90 & 0.95 & 0.99\\\hline
$(\frac{1}{2},\frac{1}{2})$   & -2.424 & -1.616 & -1.212 & 1.212 & 1.616 & 2.222\\
$(\frac{1}{16},\frac{1}{16})$ & -1.266 & -1.266 & -1.266 & 1.320 & 2.182 & 3.044\\
 \hline
\end{tabular}\\

\vspace{0.5cm}

The local correlations can be used to test independence and to verify local negative and positive dependence, as well. In particular, observe that except five values close to the vertices (0,0) and (1,1), where the standardized empirical local correlations are in [-1.212,-2.000), the remaining ones are strictly less than -2.000. This, along with the rough information contained in Table 2, allow one to expect that all local correlations (in the grid points $\mathbbm {G}_{16}$) shall be accepted to be non positive on the standard level $\alpha$=0.05.

Obviously, the local correlations can be also used to form a new test statistic on global negative dependence. A reasonable candidate is the test rejecting such hypothesis for large values of $L^*$, see (12). By Theorem 1 of Ledwina and Wy{\l}upek (2014) such statistic preserves the correlation order. For the data under consideration, simulated in 10 000 MC runs, $p$-value of this test is equal to 1.

Our conclusion is that the heat map displayed in Figure 4 of this paper supports more simple picture of the overall dependence structure than this one presented in Figure 4 of Jones and Koch (2003) and obtained via kernel methods applied to the original data.\\

The second sample of automobile data consists of  observations of acceleration time ($X$) and fuel consumption ($Y$). Both, the original data display in Jones and Koch (2003) and the scatter plot provided in Figure 5 of our paper, suggest some not very strong positive dependence. The plot of $Q_n$ supports this suggestion.
The heat map visualizes standardized correlations and gives better insight into the strength of this dependence. The strongest local correlations are observed close to the lower tails of both (transformed) variables and the strength of the dependence is getting weaker towards the upper tails. Again our look at the data reveals a simpler structure of the dependence that this one provided in Figure 5 of Jones and Koch (2003). In particular, we do not notice zero local dependence between moderately large values of both variables. Test rejecting positive local dependence for small values of $L_n(i/16,j/16), i,j=1,\ldots,15,$ can be applied in each grid point while global positive dependence can be verified by the test rejecting it for small values of $L_*$, see (12) and (11). For the given data, the simulated, on the basis of 10 000 observations, $p$-value of such global test on positive quadrant dependence is equal to  1.\\

\hspace{40 mm} { (Acceleration Time, Fuel Consumption)}, $n = 392$\\

\vspace{1 mm}

\noindent
\includegraphics[width = 4 cm, height = 3.97 cm]{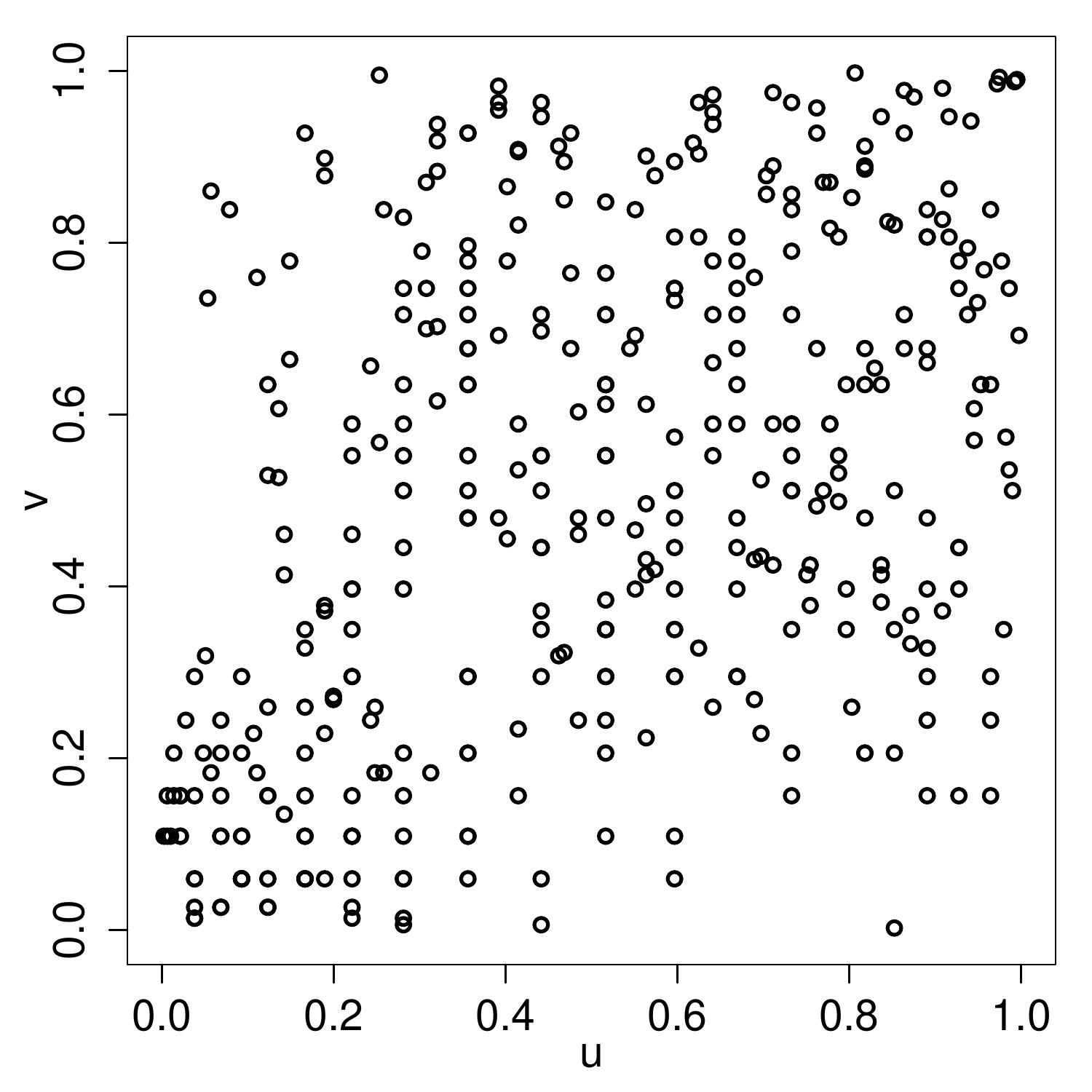}\quad \quad
\includegraphics[width = 4 cm, height = 3.97 cm]{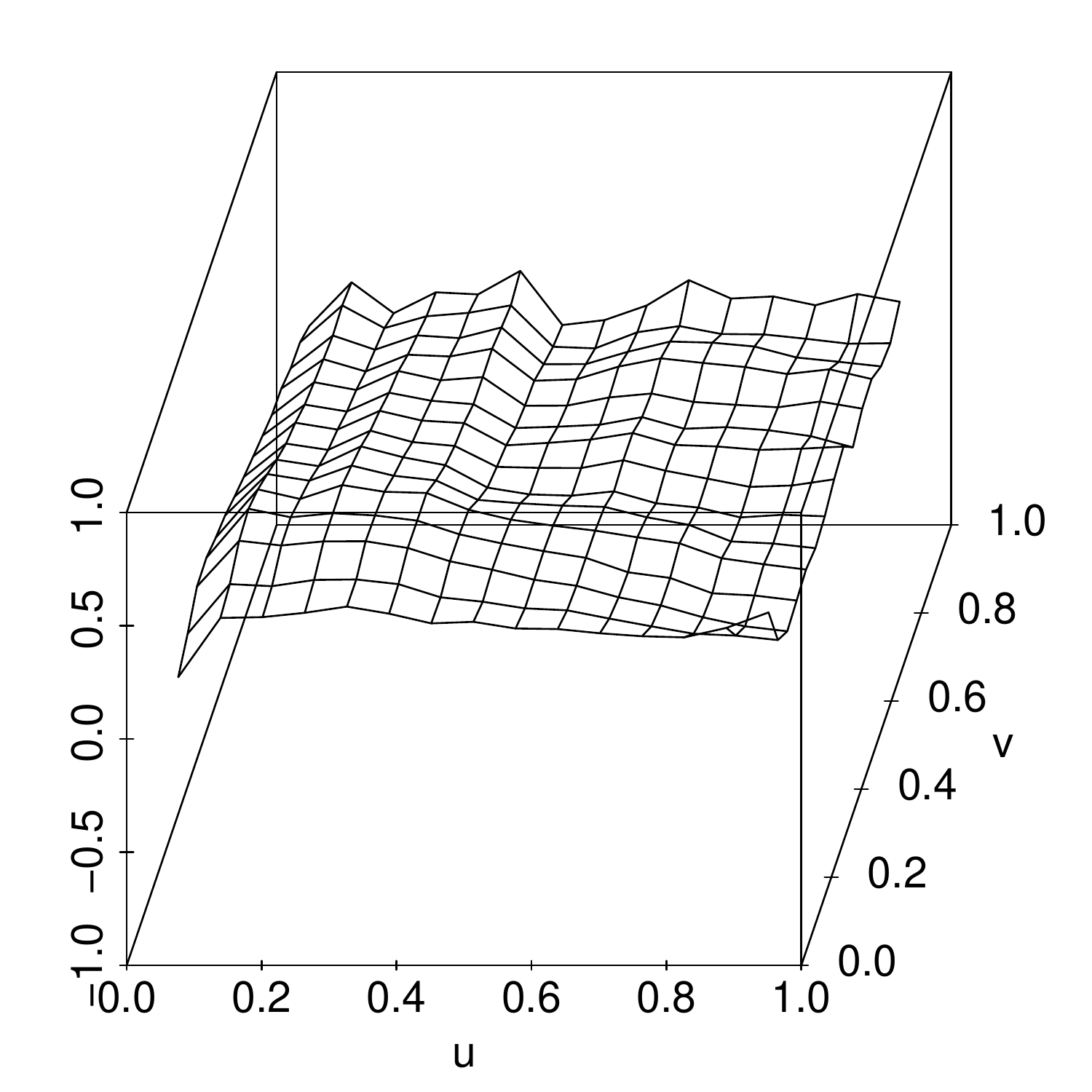}\quad
\includegraphics[width = 4 cm, height = 3.97 cm]{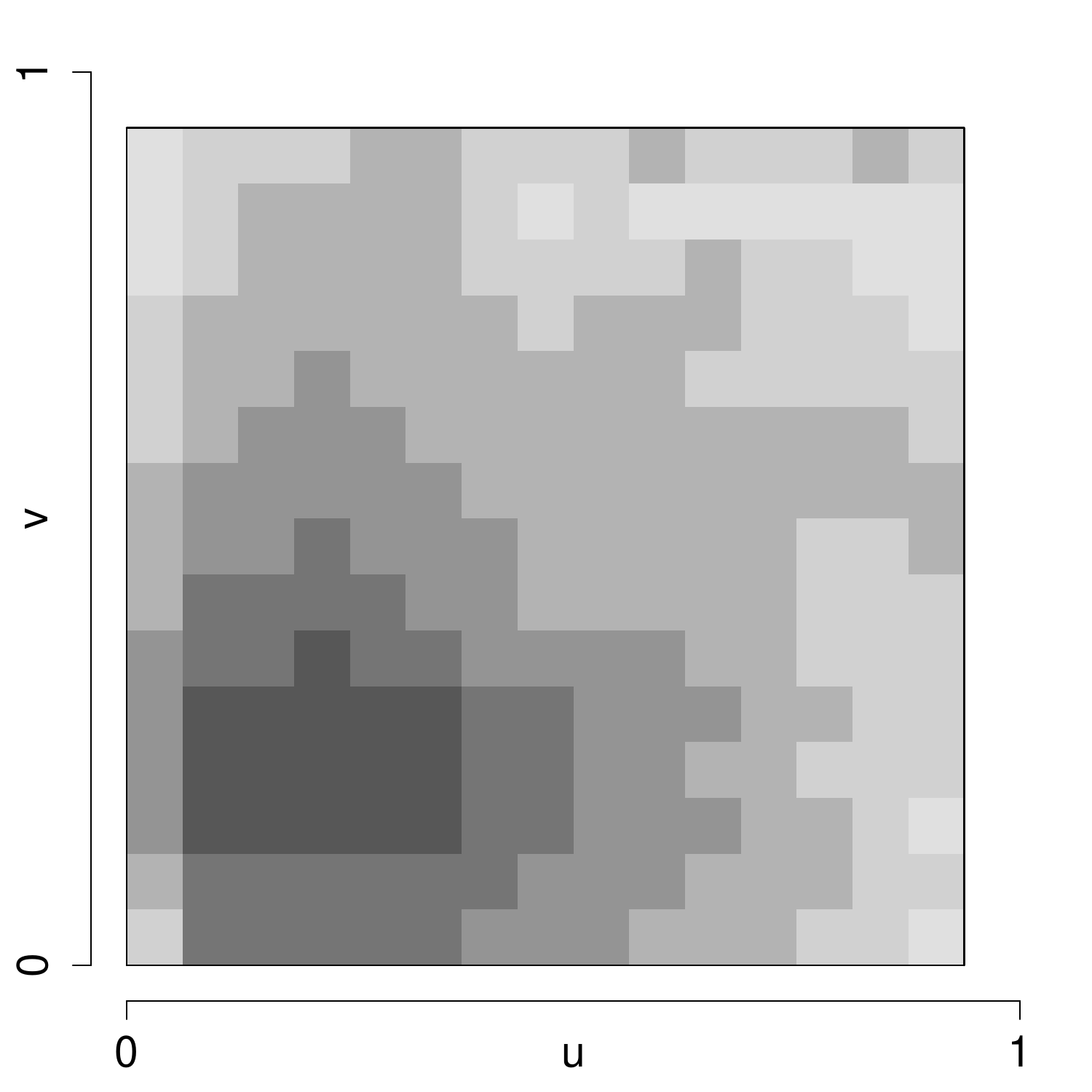}
\includegraphics[width = 4 cm, height = 3.97 cm]{Legend.pdf}

\noindent
{\it Fig. 5.} Left panel: scatter plot of \,$(R_i/(n+1),S_i/(n+1))$,\, $i = 1,\ldots,n$, $n = 392$;
middle panel: estimator $Q_n(u,v)$ of $q_C(u, v)$ on the grid $\mathbbm {G}_{16}$;
right panel: standardized estimator $L_n(u,v) = \sqrt{n}\, Q_n(u,v)$ on the grid $\mathbbm {G}_{16}$.
$L_{*} = 0.8$, $L^{*} = 10.2$.\\

\noindent
{\sl 4.3. Example 3: Aircraft data}\\

Consider $n=230$ aircraft span and speed data, on log scales, from years 1956-1984, reported and analyzed in Bowman and Azzalini (1997). We summarize the data in Figure 6. Since in this example both negative and positive correlations appear, we added respective signs to the colors in the heat maps. The figure exhibits that small and moderately large values of log speed are positively correlated with log span, while for the remaining cases the relation is reversed. Two, approximately symmetrically located, regions of relatively strong dependence are seen. In general, the strength of dependence is weaker than in previous cases. Similarly as in the previous example, also here our approach provides simpler and more regular picture of the dependence structure than this one presented in Figure 2 of Jones and Koch (2003). \\

\hspace{40 mm} { ($\log$(Span), $\log$(Speed))}, $n = 230$\\

\vspace{1 mm}

\noindent
\includegraphics[width = 4 cm, height = 3.97 cm]{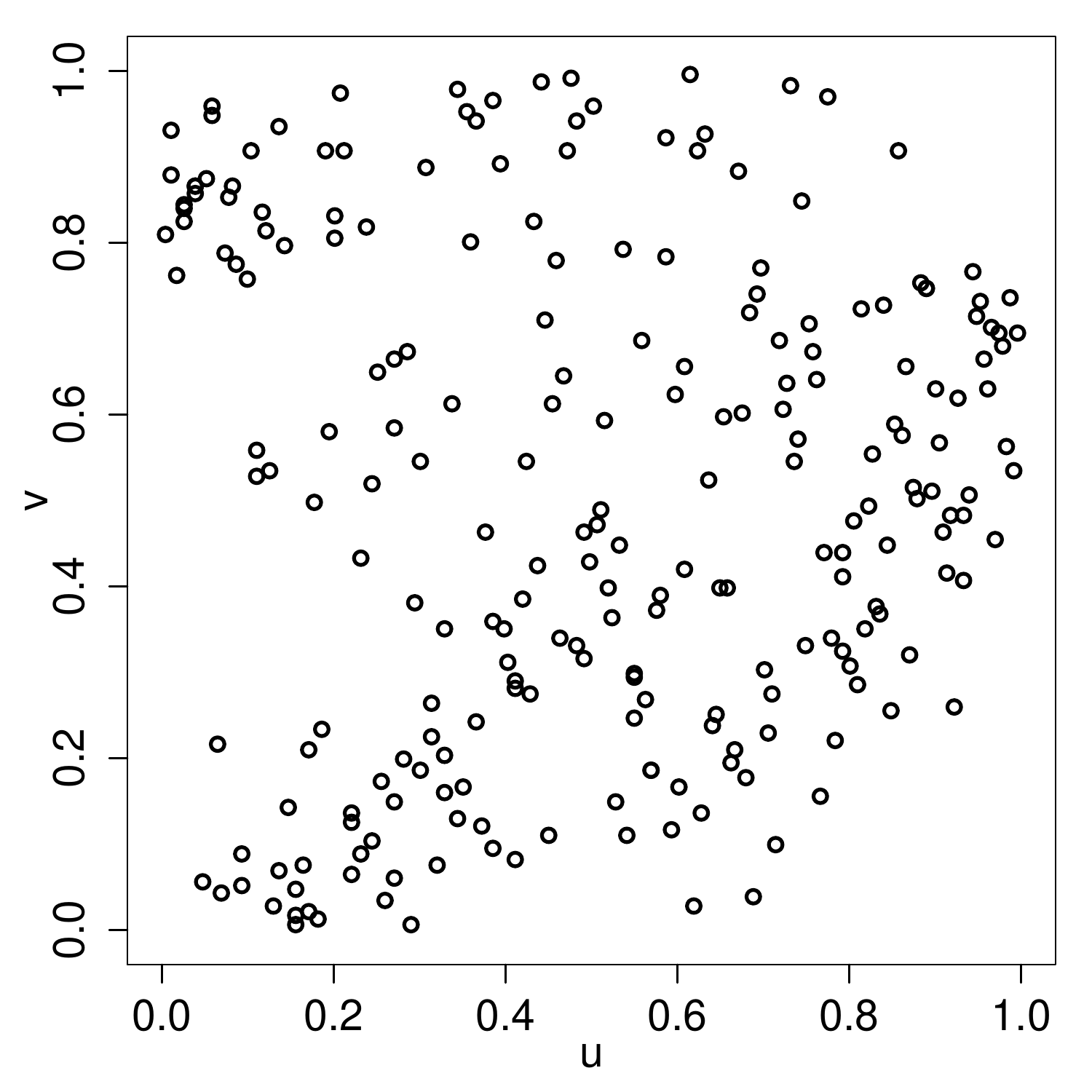}\quad \quad
\includegraphics[width = 4 cm, height = 3.97 cm]{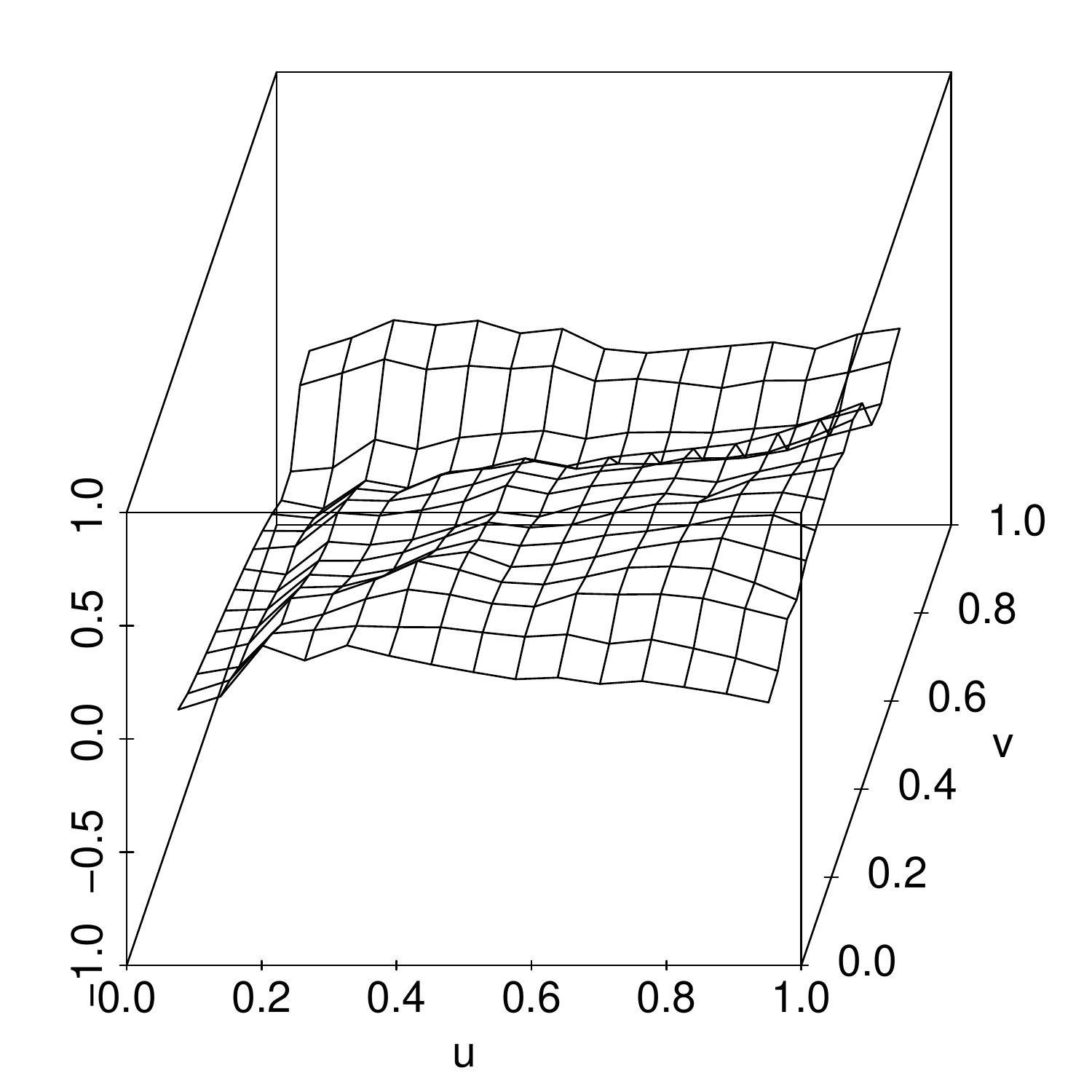}\quad
\includegraphics[width = 4 cm, height = 3.97 cm]{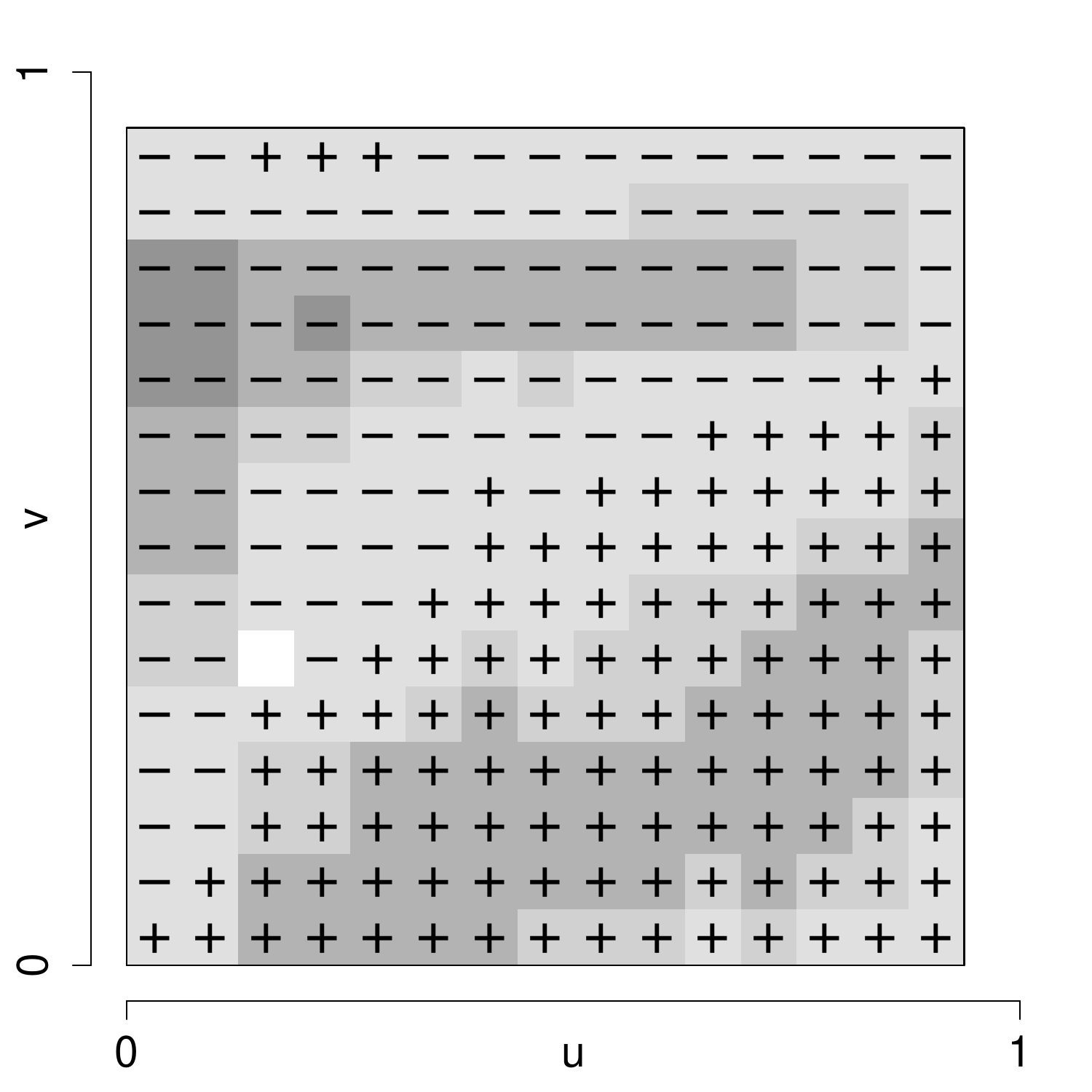}
\includegraphics[width = 4 cm, height = 3.97 cm]{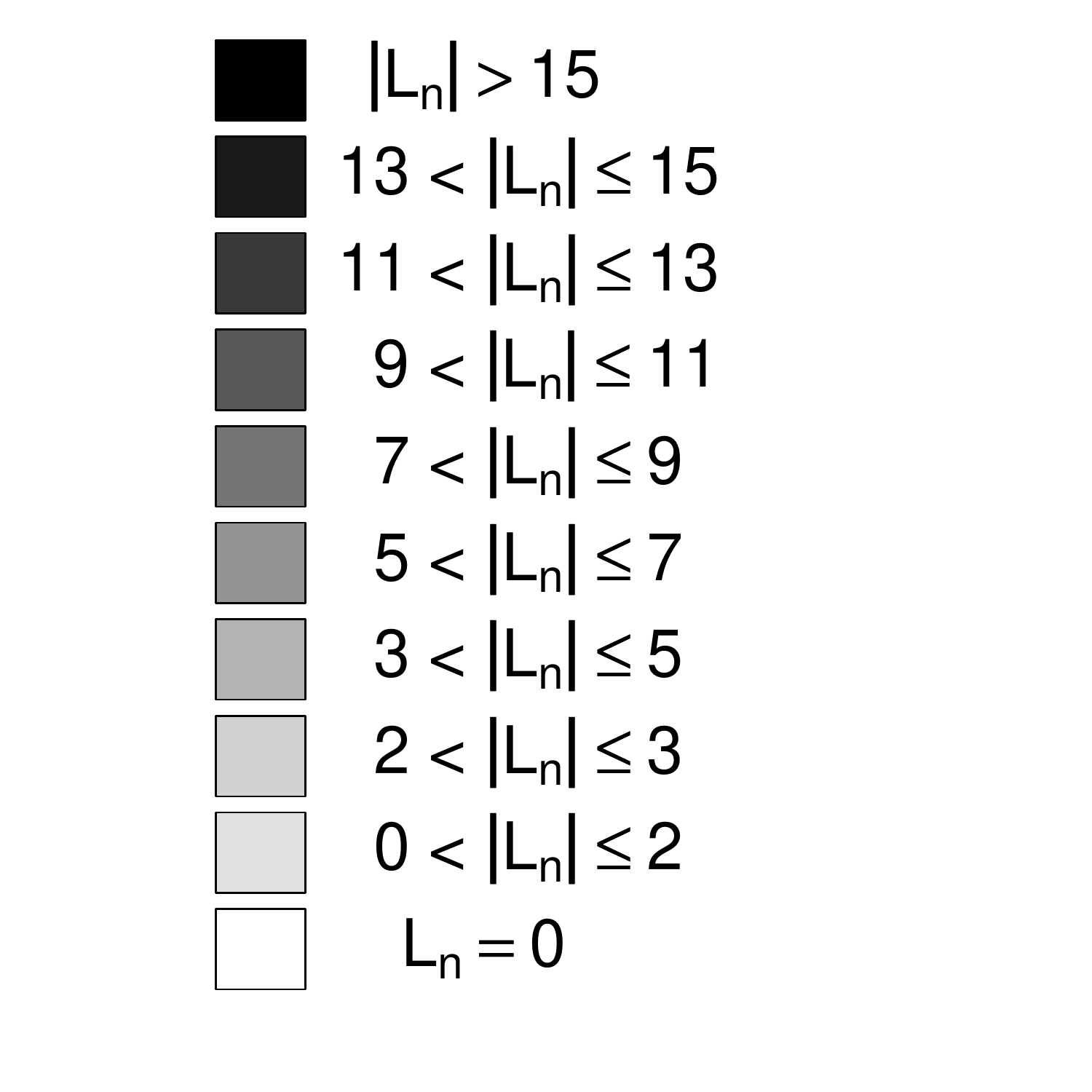}

\noindent
{\it Fig. 6.} Left panel: scatter plot of \,$(R_i/(n+1),S_i/(n+1))$,\, $i = 1,\ldots,n$, $n = 230$;
middle panel: estimator $Q_n(u,v)$ of $q_C(u, v)$ on the grid $\mathbbm {G}_{16}$;
right panel: standardized estimator $L_n(u,v) = \sqrt{n}\, Q_n(u,v)$ on the grid $\mathbbm {G}_{16}$.
$L_{*} = -6.5$, $L^{*} = 4.6$.\\

Bowman and Azzalini (1997) used these data to discuss some drawbacks of standard correlation measures. Indeed, for these data classical Pearson's, and Spearman's and Blomqvist's rank statistics for assessing an association yield simulated $p$-values 0.81, 0.74, and 0.79, respectively. Kendall's rank correlation gives simulated $p$-value 0.31, which also seems to be too high, when one is looking at the magnitude of standardized local correlations in Figure 6. Combining the local correlation into global statistic $L^o =\max_{1 \leq i,j \leq 15} |L_n(i/16,j/16)|$, with large values being significant, basing on simulation of size 10 000, we get $p$-value 0.00 for such global independence test. This shows  that local correlations are more informative than each of the above single classical global indices of association.\\

\noindent
{\bf 5. Discussion}\\

We have introduced the novel function valued measure of dependence of two random variables. Its definition, based on Studentized difference of two cdf's, is general, simple, and natural. In our considerations, we mainly focus attention on copula-based variant of the measure. It allows for simple estimation and guarantees appealing finite sample properties of the resulting estimate. The estimate is tightly linked to the popular scatter plot and helps to extract explicit dependence structure from it. Both, the measure and the estimate, allow for comparison and visualization of different association structures. The value of the measure in a fixed point has useful interpretation as correlation coefficient of some specific increasing functions of the marginals. Also, the proposed estimate features simple interpretation and easy implementation. Its performance in real data analysis yields relatively simple, in comparison to alternative method, dependence structure. We believe that the proposed approach will be useful in practice. 

Also, simple and reliable tests for local and global association, based on estimated dependencies have been proposed. It is worth  noticing that, in particular, statistic like $L^*$, cf. (12), can be considered as a usable approximation of empirical isotonic canonical correlation coefficient, introduced in Schriever (1987). Similarly, $L^o$, given in (13), can be serve as an easy to implement approximate exemplification of R\'enyi's idea to calculate maximal correlation over large class of functions. In Ledwina and Wy{\l}upek (2014) empirical correlations close to $L_n(u,v)$'s were successfully applied to construct highly sensitive test for detection of positive quadrant dependence. Recently, there is much of interest in detecting dependencies in some conditional copulas; see Veraverbeke et al. (2011), and Li et al. (2014) for discussion and further references. It seems that some graphical presentation of dependence structure of two random variables, conditionally upon  some fixed values of a covariate,  and formal application of pertaining counterparts of $L^o, L^*$,  and $L_*$ could be useful in such considerations, as well. It is also worthy  noting that the definition of $q$ can be naturally extended to higher dimensions and applied to construct tests for positive orthant dependence, for example. These questions are however beyond the scope of this initial article.\\

\noindent
{\bf References}\\

I. Bairamov, S. Kotz, T.J. Kozubowski (2003),  A new measure of linear local dependence, {\it Statistics}, { 37}, 243-258. 

N. Balakrishnan, Ch.-D. Lai (2009),  {\it Continuous Bivariate Distributions}, Springer, Dordrecht.

G. Bertensen, B. St{\o}ve, D. Tj{\o}stheim, T. Nordb{\o} (2013)  Recognizing and visualizing copulas: an approach using local Gaussian approximation, manuscript, Department of Mathematics, University of Bergen,  http://folk.uib.no/gbe062/localgaussian-correlation.

S. Bjerve, K. Doksum (1993), Correlation curves: measures of association as function of covariate values, {\it The Annals of Statistics}, { 21}, 890-902.

A.W. Bowman, A. Azzalini (1997), {\it Applied Smoothing Techniques for Data Analysis}, Clarendon Press, Oxford.

S. Cambanis, G. Simons, W. Stout (1976),  Inequalities for $Ek(X,Y)$when marginals are fixed, {\it Zeitschrift f\"{u}r Wahrscheinlichkeitstheorie und vervandte Gebiete}, { 36},  285-294.

P. Deheuvels,  La Fonction de dep\'endance empirique et ses propri\'et\'s (1979), {\it Academie Royale Belgique, Bulletin de la Classe des Sciences, 5e S\'erie}, {65}, 274-292.

J. Dhaene, M. Denuit, S. Vanduffel (2009),  Correlation order, merging and diversification, {\it Insurance: Mathematics and Economics}, { 45}, 325-332.

D. Drouet Mari, S. Kotz (2001), {\it Correlation and Dependence}, Imperial College Press, London.

P. Embrechts, A. McNeil, D. Straumann (2002),  Correlation and dependency in risk management: properties and pitfalls, in: Dempster, M., Moffatt, H. (Eds.), {\it Risk Management: Value at Risk and Beyond}, Cambridge: Cambridge University Press, Cambridge, pp. 176-223.

J.-D. Fermanian, D. Radulovi\'c, M. Wegkamp (2004),  Weak convergence of empirical copula process, {\it Bernoulli}, { 10}, 847-860.

N.I. Fisher, P. Switzer (1985), Chi-plots for assessing dependence, {\it Biometrika}, 72, 253-65.

D.M. Hawkins, Fitting monotonic polynomials to data (1994), {\it Computational Statistics}, 9, 233-247.

P.W. Holland, Y.J. Wang (1987), Dependence function for continuous bivariate densities, {\it Communications in  Statistics - Theory and Methods}, { 16}, 863-876.

R.J. Hyndman, Y. Fan (1996), Sample quantiles in statistical packages, {\it The American Statistician},  50, 361-365.

P. Janssen, J. Swanepoel, N. Veraverbeke (2012), Large sample behavior of the Bernstein copula estimator, {\it Journal of Statistical Planning and Inference}, {142}, 1189-1197.

K. Jogdeo (1982). Dependence, concepts of, in: S. Kotz, N.L. Johnson (Eds), {\it Encyclopedia of Statistical Sciences}, Vol. 2. Wiley, New York, pp. 324-334.

M.C. Jones (1998),  Constant local dependence, {\it Journal of  Multivariate Analysis}, { 64}, 148-155.

M.C. Jones, I. Koch (2003),  Dependence maps: local dependence in practice, {\it Statistics and Computing}, { 13}, 241-255.

T. Kowalczyk, E. Pleszczy\'nska (1997),  Monotonic dependence functions of bivariate distributions, {\it The Annals of Statistics}, { 5}, 1221-1227.

H.O. Lancaster (1982),  Dependence, measures and indices of, in: S. Kotz, N.L. Johnson (Eds), {\it Encyclopedia of Statistical Sciences}, Vol. 2. Wiley, New York,  pp. 334-339.

T. Ledwina, G. Wy{\l}upek (2014),  Validation of positive quadrant dependence, {\it Insurance: Mathematics and Economics}, 56, 38-47.

R. Li, Y. Cheng, J.P. Fine (2014), Quantile association regression models, {\it Journal of the American Statistical Association}, 109, 230-242.

J.-F. Mai, M. Scherer (2011),  Bivariate extreme-value copulas with discrete Pikands dependence measure, {\it Extremes}, { 14}, 311-324.

R.B. Nelsen (2006), {\it An Introduction to Copulas}, Springer, New York.

M. Omelka, I. Gijbels, N. Veraverbeke (2009), Improved kernel estimation of copulas: weak convergence and goodness-of-fit, {\it The Annals of Statistics} {37}, 3023-3058.

B. P\'oczos, Z. Ghahramani, J. Schneider (2012), Copula-based kernel dependency measures, in {\it Proceedings of the 29th International Conference on Machine Learning}, New York; Omnipress, pp. 775-782.

B. Schweizer, E.F. Wolff (1981),  On nonparametric measures of dependence for random variables, {\it The Annals of Statistics}, {9}, 879-885.

B.F. Schriever (1987), An ordering for positive dependence, {\it The Annals of Statistics}, 15, 1208-1214.

J.W.H. Swanepoel, J.S. Allison (2013),  Some new results on the empirical copula estimator with applications, {\it Statistics and Probability Letters}, { 83}, 1731-1739.

D. Tj{\o}stheim, K.O. Hufthammer (2013),   Local Gaussian correlation: A new measure of dependence, {\it Journal of Econometrics}, {172}, 33-48.

N. Veraverbeke, M. Omelka, I. Gijbels (2011), Estimation of a conditional copula and association measures, {\it Scandinavian Journal of Statistics}, 38, 766-780.

\end{document}